\documentclass[draft]{agujournal2019}
\usepackage{url} 
\usepackage{lineno}
\usepackage[inline]{trackchanges} 
\usepackage{amsmath}
\usepackage{soul}
\makeatletter
\renewcommand\paragraph{\@startsection{paragraph}{4}{\z@}%
            {-2.5ex\@plus -1ex \@minus -.25ex}%
            {1.25ex \@plus .25ex}%
            {\normalfont\normalsize\bfseries}}
\makeatother
\draftfalse

\journalname{JGR: Atmospheres}

\begin{document}

\newcommand{\eps}{\varepsilon}

\def\ex{\mathbf{e}_{x}}
\def\ey{\mathbf{e}_{y}}
\def\ez{\mathbf{e}_{z}}

\def\ephi{\mathbf{e}_{\phi}}
\def\er{\mathbf{e}_{r}}

\def\cv{\mathbf{c}}
\def\cg{\mathbf{c}_g}
\def\cgb{\mathbf{c}_{g,\beta}}
\def\cgh{\mathbf{\hat{c}}_g}
\def\cgmh{\mathbf{\hat{c}}_\gamma}
\def\cgmb{\mathbf{\hat{c}}_{\beta,\gamma}}
\def\cph{\mathbf{c}_p}
\def\Fv{\mathbf{F}}
\def\Gv{\mathbf{G}}
\def\kv{\mathbf{k}}
\def\kvb{\mathbf{k}_\beta}
\def\kvhb{\mathbf{k}_{h,\beta}}
\def\khb{k_{h,\beta}}
\def\kvp{\dot{\mathbf{k}}}
\def\pvh{\mathbf{p}_h}
\def\Rv{\mathbf{R}}
\def\sigv{\mathbf{\sigma}}
\def\tv{\mathbf{\mathcal{\tau}}}
\def\uv{\mathbf{u}}
\def\vv{\mathbf{v}}
\def\wv{\mathbf{w}}
\def\xv{\mathbf{x}}
\def\Xv{\mathbf{X}}
\def\Vv{\mathbf{V}}
\def\Uv{\mathbf{U}}
\def\Fv{\mathbf{F}}
\def\Tv{\mathbf{T}}
\def\Mv{\mathbf{M}}

\def\bh{\hat{b}}
\def\Ph{\hat{P}}
\def\uvh{\hat{\uv}}
\def\vvh{\hat{\vv}}
\def\wh{\hat{w}}

\def\bt{\tilde{b}}
\def\Pt{\tilde{P}}

\def\bb{\bar{b}}
\def\Pb{\bar{P}}
\def\rhob{\bar{\rho}}
\def\thetab{\bar{\theta}}
\def\ub{\bar{u}}
\def\vb{\bar{v}}
\def\wb{\bar{w}}

\def\vrm{\vb^*}
\def\wrm{\wb^*}

\def\ba{\langle b \rangle}
\def\bta{\langle \bt \rangle}
\def\Pta{\langle \Pt \rangle}
\def\sigva{\langle \sigv \rangle}
\def\vva{\langle \vv \rangle}
\def\uva{\langle \uv \rangle}
\def\ua{\langle u \rangle}
\def\vma{\langle v \rangle}
\def\wa{\langle w \rangle}
\def\dta{\langle \delta\theta \rangle}
\def\pia{\langle \pi \rangle}

\newcommand{\rhobar}{\overline{\rho}}

\def\tbar{\overline{\theta}}

\def\cgh{\mathbf{\hat{c}}_g}

\def\Ac{\mathcal{A}}
\def\Nc{\mathcal{N}}

\def\Gc{\mathcal{G}}
\def\Hc{\mathcal{H}}

\newcommand{\vFc}{{\boldsymbol{\mathcal{F}}}}
\newcommand{\vGc}{{\boldsymbol{\mathcal{G}}}}
\newcommand{\vHc}{{\boldsymbol{\mathcal{H}}}}

\def\oh{\hat{\omega}}

\def\Rhobarnull{{\overline{R}^{(0)}}}

\newcommand{\Pbarnull}{\overline{P}^{(0)}}

\newcommand{\pibarnull}{{\overline{\Pi}^{(0)}}}

\def\tbarnull{{\overline{\Theta}^{(0)}}}
\def\tbaralpha{\overline{\Theta}^{(\alpha)}}

\def\Vv{\mathbf{V}}
\def\Uv{\mathbf{U}}

\newcommand{\Ord}{\mathcal{O}}

\newcounter{mycount}
\newcounter{mycount2}

\newcommand{\ds}{\ensuremath\displaystyle}

\newcommand{\myDiv}[1]{\nabla\cdot#1}

\newcommand{\mySum}[3]{
  \sum\limits_{#1=#2}^{#3}
}

\newcommand{\fld}[3]{{#1}_{#2}^{(#3)}}
\newcommand{\Vfld}[2]{\Vv_{#1}^{(#2)}}
\newcommand{\Ufld}[2]{\Uv_{#1}^{(#2)}}
\newcommand{\Wfld}[2]{W_{#1}^{(#2)}}
%
%


\title{The Impact of Non-Orographic Gravity Waves on Transport and Mixing: \\
Effects of Oblique Propagation and Coupling to Turbulence}

%
%




\authors{T. Banerjee\affil{1}, S. Borchert\affil{2}, Y.-H. Kim\affil{3}, A. Kosareva\affil{1}, D. Kunkel\affil{4}, G.T. Masur\affil{1}, Z. Prochazkova\affil{5}, J. Schmidli\affil{1}, G.S. Voelker\affil{1,6}, and U. Achatz\affil{1}}


\affiliation{1}{Institut für Atmosphäre und Umwelt, Goethe Universität Frankfurt, Germany}
\affiliation{2}{Deutscher Wetterdienst, Offenbach, Germany}
\affiliation{3}{Seoul National University, Seoul, Korea}
\affiliation{4}{Institut für Physik der Atmosphäre, Universität Mainz, Germany}
\affiliation{5}{Department of Atmospheric Physics, Faculty of Mathematics and Physics, Charles University, Prague, Czech Republic}
\affiliation{6}{Leibniz-Institut für Ostseeforschung, Rostock, Germany}




\correspondingauthor{T. Banerjee}{banerjee@iau.uni-frankfurt.de}



\begin{keypoints}

\item Oblique propagation of gravity waves (GWs) cools and lowers the summer mesopause while reducing vertical shear and intensifying turbulence.

\item GW - turbulence coupling meanwhile warms and lifts the summer mesopause, but shear is still reduced and turbulence further intensified.

\item GW - turbulence coupling also enhances tracer mixing where turbulence is strongest.


\end{keypoints}

%
%

%
%


\begin{abstract}
Gravity waves (GWs) are a fundamental driver of circulation, tracer transport, and mixing in the middle and upper atmosphere, but their treatment in global circulation models remains incomplete. In particular, standard parameterizations typically restrict propagation to the vertical and treat GW–turbulence interactions in only a rudimentary manner, potentially leading to systematic biases in simulated dynamics and transport. Here, we use the Multi-Scale Gravity-Wave Model (MS-GWaM) implemented in the Community Climate Icosahedral Nonhydrostatic Model UA-ICON, together with a novel theoretical framework, to quantify the impact of (i) oblique GW propagation and (ii) explicit bidirectional coupling between GWs and turbulence. Ensemble simulations reveal that allowing for oblique propagation lowers and cools the summer mesopause by shifting the deposition of momentum and heat to lower altitudes, reduces GW-induced vertical shear in the the middle and lower atmosphere, and enhances turbulent kinetic energy in the upper mesosphere and lower thermosphere. In contrast, coupling GWs to turbulence produces a nearly opposite mesopause response, lifting and warming the mesopause, while maintaining a reduction in wave-induced shear and further enhancing turbulence. Tracer experiments additionally show that turbulent coupling significantly increases mixing, particularly in regions of enhanced TKE, with implications for chemical redistribution. These results demonstrate that both oblique GW propagation and GW–turbulence interactions exert leading-order controls on mesosphere–lower thermosphere circulation, temperature structure, and tracer transport. Neglecting these processes in global models likely contributes to biases in the Brewer–Dobson circulation, energy balance, and constituent distributions, underscoring the need for next-generation GW parameterizations that capture these effects.
\end{abstract}

\section*{Plain Language Summary}

In the middle and upper atmosphere, small-scale waves called gravity waves (GW) caused by events like thunderstorms or jet streams strongly influence winds, temperature, and circulation. This paper explores how non-orographic GWs, especially those traveling at an angle to the vertical (obliquely), alter transport and mixing of air. Unlike vertically propagating GWs, oblique GWs can deposit their momentum and energy across a wider region and altitudes.

Using a state-of-the-art GW parameterization and atmospheric model alongside a novel theoretical framework, we simulate these oblique GWs and their interaction with turbulence - the chaotic motions of the atmosphere. Results show that neglecting these processes may lead models to underestimate transport, mixing, and circulation, particularly in the tropical stratosphere and mesosphere where GW activity is strongest.

Oblique propagation is found to cool and lower the summer mesopause (the coldest region of the atmosphere), while reducing vertical shear and intensifying turbulence. In contrast, when GWs are coupled to turbulence, the mesopause warms and rises, but shear is still reduced and turbulence intensified. This coupling also enhances mixing, influencing heat and chemical redistribution.

Overall, the findings point to the need for GW parameterizations that account for oblique propagation and turbulence coupling.

\section{Introduction}

The transport of tracers through the middle atmosphere, with its implications for the local heat budget, radiation, and chemistry, is a key issue of climate science. Two complementary concepts have proven to be essential for the understanding of atmospheric tracer transport \cite{cohenWhatDrivesBrewer2014} in the Brewer-Dobson circulation (BDC). The first is advection under the influence of Rossby waves (RW) and gravity waves (GW). The corresponding residual circulation is driven by RW and GW breaking and their impact on the resolved momentum. The second is the impact of irreversible mixing due to turbulence, among other processes resulting from RW and GW breaking as well. Hence turbulence and GWs feature as non-negligible factors for tracer transport through the upper troposphere / lower stratosphere (UTLS) and above. Neither of the two are fully resolved by weather-forecast and climate models, so that their effect on tracers has to be taken into account by appropriate parameterizations. Yet, the direct impact of GWs on tracers and on turbulence is typically neglected, while studies indicate that climate models have difficulties to accurately model the transport processes in the UTLS, 
e.g.,
being very sensitive to model resolution and choice of sub-grid scale parameterizations \cite<e.g.>{hegglin_multimodel_2010}.

Present-day handling of the GW impact in climate models still suffers from various limitations \cite{achatz_atmospheric_2024}. (i) Among others, GW parameterizations are typically neglecting both the impact of transient GW interactions with the resolved flow and the fact of oblique GW propagation as well as the impact of horizontal variations of the resolved flow on GWs. A more general treatment has by now become possible,
using the Lagrangian parameterization approach of the Multi-Scale Gravity-Wave Model MS-GWaM \cite{muraschko_application_2015,boloni_interaction_2016,achatz_multi-scale_2023}.
Corresponding studies \cite{boloni_toward_2021,kim_toward_2021,kim_crucial_2024,voelker_ms-gwam_2024} show that it has a significant effect on zonal-mean winds and temperature, including the quasi-biennial oscillation. A related effect on the residual circulation is to be expected, but this has not been studied yet. (ii) Moreover, although there is a vigorous interaction between GWs and turbulence, e.g. during GW breaking \cite<e.g.>{fritts_gravity_2003,achatz_gravity-wave_2007,fritts_gravity_2009-1,fritts_gravity_2009}, it is only represented in a comparatively rudimentary way in existing parameterizations. Typically the saturation approach of \citeA{lindzen_turbulence_1981} is used to reflect the damping of GWs by turbulence due to their breaking, but a corresponding source is not considered in the turbulence parameterization. Turbulent diffusion also does not act on parameterized GWs, even if the source of this turbulence is other GWs.
Corresponding effects on turbulent tracer mixing have also not been studied yet, to the best of our knowledge, and they are not captured in operational GW and turbulence parameterizations. (iii) Finally, tracer fluxes due to subgrid-scale gravity waves, related to the Stokes drift \cite<e.g.>{buhler_waves_2009} are presently also not represented in the transport of tracers in atmospheric models.

The present study sets out to address issues (i) and (ii). With this purpose, Section \ref{sec:theory} lays forth the necessary theoretical groundwork while Section \ref{sec:experiment} provides results from ensemble simulations. Finally, Section \ref{sec:summary} then summarizes the key findings and discusses future potential developments. Additionally, also provided in the Appendix is the detailed implementation of the developed theory into 
MS-GWaM in the community climate and weather prediction code UA-ICON \cite{borchert_upper-atmosphere_2019,banerjee_2025_code}.

\section{Theory}\label{sec:theory}

This section is divided into seven parts. Firstly, it presents a derivation of the residual mean circulation in spherical coordinates, incorporating an extended Stokes drift formulation that accounts for both the resolved and unresolved waves. The associated downward-control framework is then developed within the same context. Furthermore, the bidirectional coupling between the gravity wave parameterization and the turbulence scheme is derived in detail, elucidating the mechanisms on both sides of the interaction. The section also addresses the effects of oblique wave propagation and clarifies the use of the pseudomomentum approximation employed throughout this study. It then ends with the model description and setup used in the following section on experimentation.

\subsection{Residual-Mean Circulation}\label{RMC}

Beginning with the residual-mean circulation, to diagnose the zonal-mean tracer transport without mixing, the transformed Eulerian mean from \citeA{andrews_generalized_1978} can be used to define residual-mean meridional and vertical winds in spherical coordinates ($\lambda,\phi,r$) with longitude $\lambda$, latitude $\phi$, and radial distance $r$ from the center of earth,
\begin{eqnarray}
    \vrm (\phi,r)
    &=& \frac{1}{\rhob} \left(\overline{\rho v} - \frac{1}{r \cos\phi} \partial_r \Psi_s\right)
    = - \frac{1}{\rhob} \frac{1}{r \cos\phi} \partial_r \left(\Psi_e + \Psi_s\right)
    \label{eq_vrm}\\
    \wrm (\phi,r)
    &=& 
    \frac{1}{\rhob} \left(\overline{\rho w} + \frac{1}{r^2 \cos\phi} \partial_\phi \Psi_s\right)
    = \frac{1}{\rhob} \frac{1}{r^2 \cos\phi} \partial_\phi \left(\Psi_e + \Psi_s\right)
    \label{eq_wrm}
\end{eqnarray}
Here the overbar indicates a zonal mean, ($\rho$ is the density) and the residual mean streamfunction $\Psi = \Psi_e + \Psi_s$ has two parts, the Eulerian-mean mass stream function $\Psi_e$, and the Stokes drift contribution $\Psi_s$.
Under the condition that $\partial_t \overline{\rho}$ is negligibly small in magnitude, the zonal-mean continuity equation,
\begin{equation}
    \partial_t \overline{\rho} + \nabla\cdot\left(\overline{\rho\vv}\right) = 0
\end{equation}
implies that $\Psi_e$ satisfies,
\begin{equation}
    \left(\overline{\rho v},\overline{\rho w}\right) 
    = \frac{1}{r \cos\phi} \left(-\partial_
    r\Psi_e, \partial_\phi\Psi_e/r\right)
\end{equation}
 Simultaneously, the Stokes drift contribution $\Psi_s$ can be further divided as
\begin{equation}
    \Psi_s = \Psi_{rw} + \Psi_{gw}
\end{equation}
with a resolved-wave part (Rossby waves and large-scale gravity waves),
\begin{equation}
    \Psi_{rw} 
    = 
    \frac{r \cos\phi}{\left|\nabla\thetab\right|^2} 
    \left[
    \overline{\left(\rho v\right)'\theta'} \partial_r\thetab 
    - \overline{\left(\rho w\right)'\theta'} \frac{1}{r} \partial_\phi\thetab
    \right]
\end{equation}
where $\theta$ is the potential temperature, primes indicate deviations from the zonal mean, and a part due to parameterized gravity waves,
\begin{equation}
    \Psi_{gw} 
    = 
    \frac{r \cos\phi}{\left|\nabla\thetab\right|^2} 
    \overline{\left\langle\left(\rho v\right)'\theta'\right\rangle} \partial_r\thetab 
\end{equation}
where angle brackets indicate a local phase average over the small gravity-wave scales, and the primes the gravity-wave fluctuations. After having determined the residual-mean winds from (\ref{eq_vrm}) and (\ref{eq_wrm}), the residual-mean streamfunction is then obtained by integrating (\ref{eq_vrm}) assuming that $\Psi$ vanishes as $r\rightarrow\infty$ such that,
\begin{equation}\label{eq_vrm2psi}
    \Psi = \int_r^\infty dr' \rhob r'\cos\phi \,\vrm
\end{equation}

\subsection{Downward-Control Analysis}\label{DCA}

An interpretation 
and separation
of the effects of resolved waves (RW) and parameterized gravity waves (GW) on the residual-mean circulation, beyond their Stokes drift contributions, can be obtained from downward-control analysis \cite{haynes_downward_1991}. We take the zonal average of the zonal momentum equation,
\begin{equation}
    \rho \mathrm{D}_t u -\frac{\rho v u}{r} \tan\phi + \frac{\rho w u}{r} 
    - 2  \Omega \left(\rho v \sin\phi - \rho w \cos\phi\right)
    = - \frac{1}{r\cos\phi} \partial_\lambda p + \rho \left(D_{gw} + D_{rs}\right)
\end{equation}
where $\mathrm{D}_t$ is the Lagrangian derivative, $p$ the pressure, and $D_{gw}$, $D_{rs}$ the drags by unresolved gravity waves and other processes, 
respectively. Again assuming $\rhob$ to be sufficiently invariant in time and introducing the residual-mean wind from (\ref{eq_vrm}) and (\ref{eq_wrm}), we obtain
\begin{eqnarray}
    &&\partial_t \overline{u} 
    + \vrm \left[\frac{\partial_\phi \left(\ub \cos\phi\right)}{r \cos\phi} - f\right]
    + \wrm \left[\frac{\partial_r\left(r \ub\right)}{r} + \beta r\right]\nonumber\\
    &&= 
    \frac{1}{\rhob} 
    \left[
    a \frac{\nabla\cdot\left(\Fv^{rw} + \mathbf{S}^{gw}\right)}{r \cos\phi} 
    + \overline{\rho \left(D_{gw} + D_{rs}\right)} 
    - \partial_t \overline{\rho' u'}
    \right]
    \label{eq_zonmom_av}
\end{eqnarray}
where $f$ and $\beta$ are Coriolis parameter and its meridional derivative, 
$\Fv^{rw} = F^{rw}_\phi\ephi + F^{rw}_r\er$ the Eliassen-Palm flux with components
\begin{eqnarray}
    F^{rw}_\phi 
    &=& 
    - \frac{r}{a} \cos\phi \, \overline{\left(\rho v\right)'u'} 
    + \frac{\Psi_{rw}}{a} \left[\frac{\partial_r\left(r \ub\right)}{r} + \beta r\right]\\
    F^{rw}_r 
    &=& 
    - \frac{r}{a} \cos\phi \, \overline{\left(\rho w\right)'u'} 
    - \frac{\Psi_{rw}}{a} 
    \left[\frac{\partial_\phi \left(\ub \cos\phi\right)}{r \cos\phi} - f\right]
\end{eqnarray}
due to resolved waves
(mostly Rossby waves at the model resolutions applied below)
and $\mathbf{S}^{gw} = S^{rw}_\phi\ephi + S^{rw}_r\er$ with components
\begin{eqnarray}
    S^{gw}_\phi 
    &=& 
    \frac{\Psi_{gw}}{a} \left[\frac{\partial_r\left(r \ub\right)}{r} + \beta r\right]\\
    S^{gw}_r 
    &=& 
    - \frac{\Psi_{gw}}{a} 
    \left[\frac{\partial_\phi \left(\ub \cos\phi\right)}{r \cos\phi} - f\right]
\end{eqnarray}
due to subgrid-scale GW. Here $a$ is the radius of the earth.
Finally, since the sum of advection terms with $(\vrm, \wrm)$ in (\ref{eq_zonmom_av}) reduce to $\propto (d\Psi/dr)_{M=const}$ where $M$ [$\equiv r \cos\phi (\overline{u} + \Omega r \cos\phi)$] is the zonal-mean specific angular momentum, then, one can use the fact that $M=const.$ curves in the $(\phi,r)$ plane are nearly vertical \cite{haynes_downward_1991} to solve for $\vrm$ and obtain therefrom the residual-mean streamfunction using (\ref{eq_vrm}),
i.e.,
\begin{eqnarray}
    \Psi 
    &=& 
    \int_r^\infty dr' r'\cos\phi
    \frac{\ds
    C_{rw} + C_{gw} + \overline{\rho D_{rs}} 
    - \partial_t \overline{\rho' u'} - \rhob \partial_t \overline{u}
    }{\ds
    \frac{\partial_\phi \left(\ub \cos\phi\right)}{r \cos\phi} - f
    }\\
    C_{rw} &=& a \frac{\nabla\cdot\mathbf{F}^{rw}}{r \cos\phi}\\
    C_{gw} &=& \overline{\rho D_{gw}} 
    + a \frac{\nabla\cdot\mathbf{S}^{gw}}{r \cos\phi} 
    \label{eq:c_gw}
\end{eqnarray}
The streamfunction can then be de-composed into a RW part resulting only from $C_{rw}$, a GW part resulting only from $C_{gw}$, and three typically small contributions from other parameterized subgrid-scale processes and averaged time derivatives. 
Under typical conditions,
it can be shown that $C_{gw}$ agrees with the convergence of the flux of zonal GW pseudomomentum \cite<e.g.>{achatz_atmospheric_2022}. This convergence is often called GW Eliassen-Palm-flux divergence.

\subsection{Contributions of Subgrid-Scale Gravity Waves}

Using MS-GWaM \cite{achatz_multi-scale_2023,voelker_ms-gwam_2024,kim_crucial_2024}, one obtains the gravity-wave meridional Stokes drift via the wavenumber integral,
\begin{equation}
    \Psi_{gw} 
    \approx \frac{r\cos\phi}{\partial_r\thetab} \overline{\rho \langle v'\theta'\rangle} 
    \approx fr\cos\phi \overline{\int d^3 k\, \frac{k \Nc c_{gr}}{\oh^2 - f^2}}
\end{equation}
where in the integrand $\Nc$ is the spectral wave-action density, $k$ the zonal wavenumber, $c_{gz}$ the vertical group velocity and $\oh$ the intrinsic frequency. The gravity-wave-drag contribution to $C_{gw}$ is 
\begin{equation}\label{eq_gw_wave_drag}
    \overline{\rho D_{gw}}
    =
    - \overline{\nabla\cdot\rho \langle\vv' u'\rangle}
    - f \overline{\rho \langle v'\theta' \rangle / \theta}
\end{equation}
with the flux of zonal momentum having components,
\begin{eqnarray}
\label{eq_momflux_uu_dim}
    \rho \langle u' u'\rangle
    &=&
    \int d^3k \left(\frac{k \Nc \hat{c}_{g\lambda}}{1 - f^2/\oh^2}
    + \frac{l \Nc \hat{c}_{g\phi}}{\oh^2/f^2 - 1}\right) \\
    \rho \langle v' u'\rangle &=& \int d^3k\, k \Nc \hat{c}_{g\phi} = \int d^3k\, l \Nc \hat{c}_{g\lambda} \\
    \rho \langle w' u'\rangle &=& \int d^3k \frac{k \Nc \hat{c}_{gr}}{1 - f^2/\oh^2}
    \label{eq_momflux_wv_dim}
\end{eqnarray}
The pseudo-momentum approximation to this is discussed below.

\subsection{Coupling between Gravity Waves and Turbulence}\label{sec: GWT coupling}

The interaction between turbulence, gravity waves and a mean flow is approached as a multi-scale problem where turbulence lives on the smallest scales, gravity waves on intermediate mesoscales, and the mean flow on the largest synoptic scales. For a discussion of the interaction between parameterized gravity waves and parameterized turbulence, we therefore first recapitulate the derivation of the interaction between turbulence and a mean flow incorporating gravity waves. The general theory is then approximated by a flux-gradient approach for the turbulent fluxes where turbulent viscosity and diffusivity are determined from a mixing length and a characteristic turbulent velocity scale, obtained from the turbulent kinetic energy density. This leads to a system where turbulence is parameterized but gravity waves and synoptic scale flow together constitute the large-scale flow. In a final step that large scale flow and the distribution of the turbulent energy densities are divided into their mesoscale and synoptic-scale part, 
neglecting for the time being any mesoscale contributions to the turbulence distribution,
and their interaction is estimated on the basis of WKB theory. This will lead most especially to a prediction of turbulence generation from parameterized gravity waves, and also of the damping of parameterized gravity waves by turbulent viscosity and diffusivity.

The discussion is using the Boussinesq approximation for the description of atmospheric dynamics. 
This
is justified for processes on vertical scales sufficiently shorter than the atmospheric scale height. One can therefore apply it relatively safely to turbulence dynamics, but care has to be taken to incorporate the effect of the vertically decreasing density whenever turbulence is transported vertically over distances comparable to the atmospheric scale height.

\subsubsection{Energetics of the Interaction between Turbulence and Laminar Flow}

Consider the viscous-diffusive Boussinesq equations on an $f$-plane,
\begin{eqnarray}
    \left(\partial_t + \vv\cdot\nabla\right)\vv + f \ez\times\vv 
    &=& -\nabla\Pt + \bt\ez + \frac{1}{\rho_0} \nabla\cdot\sigv \label{eq_bouss_mom}\\
    \left(\partial_t + \vv\cdot\nabla\right)\bt + w N^2
    &=& \nabla\cdot\left(\kappa\nabla b\right) \label{eq_bouss_b}\\
    \nabla\cdot\vv \label{eq_bouss_cont}
    &=& 0
\end{eqnarray}
where $\vv$ is the three-dimensional wind field, $f$ the Coriolis parameter, $\ez$ the vertical unit vector, $(\Pt,\bt) = (P-\Pb,b-\bb)$ the deviations of density-normalized pressure $P$ and buoyancy $b$ from a hydrostatic reference profile $(\Pb,\bb)$ satisfying $d_z \Pb = \bb$, 
$N^2 = d_z\bb$ the reference-atmosphere stratification ($N$ being the Brunt-Väisälä frequency),
$\sigv$ the viscous stress tensor, 
\begin{equation}
    \sigv = \eta \left[\nabla\vv + \left(\nabla\vv\right)^t\right]
\end{equation}
with cartesian-coordinate elements $\sigma_{ij} = \eta \left(\partial_i v_j + \partial_j v_i\right)$, and $\eta$ and $\kappa$ the molecular (shear) viscosity and diffusivity, respectively. Inserting the Reynolds decomposition,
\begin{equation}
    \left(\vv,\bt,\Pt\right) = \left(\vva,\bta,\Pta\right) + \left(\vv',b',P'\right)
\end{equation}
into an ensemble average and deviations therefrom into the Boussinesq equations and averaging the latter yields the mean-flow equations,
\begin{eqnarray}
    \left(\partial_t + \vva\cdot\nabla\right)\vva + f \ez\times\vva 
    &=& -\nabla\Pta + \bta\ez + \frac{1}{\rho_0} \nabla\cdot\left(\sigva +\sigv_t\right) \label{eq_reymean_mom}\\
    \left(\partial_t + \vva\cdot\nabla\right)\bta + \wa N^2
    &=& \nabla\cdot\left(\kappa\nabla \ba - \langle\vv'b'\rangle\right)\label{eq_reymean_b}\\
    \nabla\cdot\vva
    &=& 0 \label{eq_reymean_cont}
\end{eqnarray}
where $\sigv_t = - \rho_0 \langle\vv'\vv'\rangle$ is the turbulent stress tensor.

\subsubsection{General Budget Equations}

Taking the scalar product of (\ref{eq_reymean_mom}) with $\vva$, of (\ref{eq_reymean_b}) with $\ba/N^2$ , using (\ref{eq_reymean_cont}), and assuming that $N^2$ is sufficiently invariant in space and time, yields the prognostic equations,
\begin{eqnarray}
    \left(\partial_t + \vva\cdot\nabla\right) K_m 
    + \nabla\cdot
    \left[
    \vva\Pta - \frac{\vva}{\rho_0} \cdot \left(\sigva - \rho_0 \langle\vv'\vv'\rangle\right)
    \right]
    &=&
    B_m - S - \epsilon_m \label{eq_kmean}\\
    \left(\partial_t + \vva\cdot\nabla\right) A_m 
    + \nabla\cdot
    \left[
    \langle\vv'b'\rangle\frac{\bt}{N^2} - \kappa\nabla A_m
    \right]
    &=&
    - B_m - C - D_m \label{eq_amean}
\end{eqnarray}
for the mass-specific mean-flow kinetic-energy density and available potential energy density
\begin{equation}
    \left(K_m,A_m\right) = \left(\frac{\left|\vva\right|^2}{2}, \frac{\bta^2}{2 N^2}\right)
\end{equation}
where $B_m = \wa\bta$ is the buoyant exchange between the two energy reservoirs, with $w$ the vertical wind, $S = -\langle\vv'\vv'\rangle \cdot\cdot \nabla\vva$ (or -$\langle v'_iv'_j\rangle\frac{\partial\langle v_i\rangle}{\partial x_j}$ in Cartesian coordinates) the turbulent shear-production rate, $C = -\langle b'\vv'\rangle \cdot\nabla\bta$ the turbulent convective-production rate, $\epsilon_m = \sigva \cdot\cdot \nabla\vva/\rho_0$ (or $\frac{1}{\rho_0}\langle\sigma_{ij}\rangle\frac{\partial\langle v_i\rangle}{\partial x_j}$ in Cartesian coordinate) the mean-flow dissipation rate, and $D_m = \kappa\left|\nabla\bta\right|^2/N^2$ the mean-flow diffusive-loss rate. 

\begin{figure}
 \centering
 \noindent\includegraphics[width=0.6\textwidth]{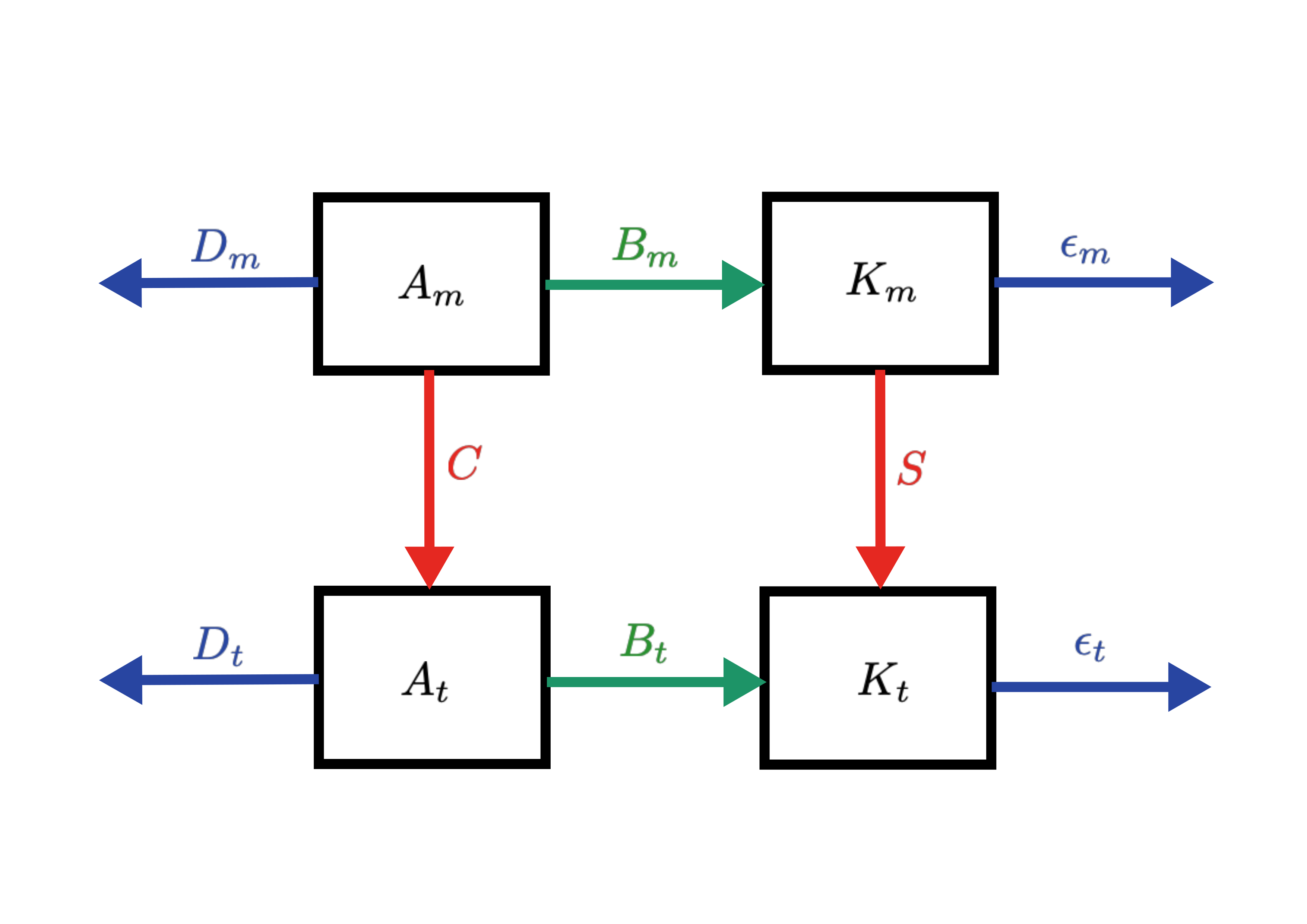}
 \vspace{-3mm}
\caption{The energy budget for interactions between turbulence and mean flow. Here $A,K$ represent the available potential and kinetic energy while $m,t$ represent the mean and turbulent regime respectively. The arrows represents the flow of energy when the respective terms are positive.}
\label{fig_budget}
\end{figure}

For the turbulence energetics first take the scalar product of (\ref{eq_bouss_mom}) with $\vv$ and multiply (\ref{eq_bouss_b}) by $\bt/N^2$ resulting in,
\begin{eqnarray}
    \left(\partial_t + \vv\cdot\nabla\right)\frac{|\vv|^2}{2} 
    &=& -\nabla\cdot \left(\vv P\right) + \bt w + \frac{1}{\rho_0} \nabla\cdot\left(\vv\sigv\right) - \frac{1}{\rho_0} \sigv\cdot\cdot\nabla\vv \label{eq_ktot}\\
    \left(\partial_t + \vv\cdot\nabla\right) \frac{\bt^2}{2 N^2} 
    &=& -w\bt + \nabla\cdot\left(\kappa \nabla\frac{\bt^2}{2 N^2}\right) -\kappa \frac{|\nabla\bt|^2}{N^2} \label{eq_atot}
\end{eqnarray}
Subtracting (\ref{eq_kmean}) -- (\ref{eq_amean}) from the average of these equations and using $\nabla\cdot\vv' = 0$ yields the prognostic equations,
\begin{eqnarray}
    \left(\partial_t + \vva\cdot\nabla\right) K_t
    + \nabla\cdot
    \left[
    \langle\vv' P'\rangle - \frac{\langle\vv'\sigv'\rangle}{\rho_0} + \frac{1}{2} \langle\vv' |\vv'|^2\rangle
    \right]
    &=&
    B_t + S - \epsilon_t \label{eq_kt}\\
    \left(\partial_t + \vva\cdot\nabla\right) A_t 
    + \nabla\cdot
    \left[
    \frac{\langle\vv'b'^2\rangle}{N^2} - \kappa\nabla A_t
    \right]
    &=&
    - B_t + C - D_t \label{eq_at}
\end{eqnarray}
for the mass-specific turbulent kinetic-energy density and available potential energy density,
\begin{equation}
    \left(K_t,A_t\right) = \left(\frac{\langle\left|\vv'\right|^2\rangle}{2}, \frac{\langle b'^2\rangle}{2 N^2}\right)
\end{equation}
where $B_t = \langle w'b'\rangle$ is the turbulent buoyant exchange between the two energy reservoirs, $\epsilon_t = \langle\sigv' \cdot\cdot \nabla\vv'\rangle/\rho_0$ the turbulent dissipation rate, and $D_t = \kappa\langle\left|\nabla b'\right|^2\rangle/N^2$ the turbulent diffusive-loss rate. The budget of the four energy reservoirs of mean flow and turbulence together with the exchange terms is illustrated in Figure \ref{fig_budget}. One sees that the generation of turbulent energy is due to shear production ($S$), transforming mean-flow kinetic energy into turbulent kinetic energy ($B_t$), and convective production ($C$) which produces turbulent available potential energy from mean-flow available potential energy.

\subsubsection{Parameterization by a TKE Approach}

The turbulence model used here \cite{doms_cosmo-model_2021} takes the standard approach of approximating the turbulent fluxes by a flux-gradient approach, i.e. it sets,
\begin{eqnarray}
    \sigma_{t,ij} &=& \eta_t \left(\partial_i \langle v_j\rangle + \partial_j \langle v_i\rangle\right)\\
    - \langle\vv' b'\rangle &=& \kappa_t \nabla\ba
\end{eqnarray}
where the turbulent viscosity $\eta_t$ and the turbulent diffusivity $\kappa_t$ are essentially given by the product of $\sqrt{K_t}$ with mixing lengths that are obtained from mean-flow properties in a rather complex way that is, however, not of major importance here. The mean-flow equations thus take the form,
\begin{eqnarray}
    \left(\partial_t + \vva\cdot\nabla\right)\vva + f \ez\times\vva 
    &=& -\nabla\Pta + \bta\ez + \frac{1}{\rho_0} \nabla\cdot\left(\sigva +\sigv_t\right)\label{eq_mean_mom}\\
    \sigva +\sigv_t &=& \left(\eta + \eta_t\right)\left[\nabla\vva + \left(\nabla\vva\right)^t\right]\\
    \left(\partial_t + \vva\cdot\nabla\right)\bta + \wa N^2
    &=& \nabla\cdot\left[\left(\kappa + \kappa_t\right)\nabla \ba \right]\label{eq_mean_b}\\
    \nabla\cdot\vva
    &=& 0 \label{eq_mean_cont}
\end{eqnarray}
The prognostic equation (\ref{eq_kt}) for the turbulent kinetic-energy density is replaced by a parameterized version,
\begin{equation}
    \left(\partial_t + \vva\cdot\nabla\right) K_t
    + \nabla\cdot\left(\tilde\eta \nabla K_t\right)
    =
    B_t + S - \epsilon_t
\end{equation}
where $\tilde\eta$ is an effective diffusion coefficient and, 
\begin{eqnarray}
    B_t &=& -\kappa_t \nabla\ba\\
    S &=& \frac{\eta_t}{\rho_0} \left[\nabla\vva + \left(\nabla\vva\right)^t\right] \cdot\cdot \nabla\vva 
    = \frac{\eta_t}{\rho_0} \left(\partial_i \langle v_j\rangle \partial_i \langle v_j\rangle + \partial_i \langle v_j\rangle \partial_j \langle v_i\rangle\right)\\
    \epsilon_t &=& K_t/\tau
\end{eqnarray}
with $\tau$ a dynamically dependent dissipation time scale. Because horizontal winds $\uva$ are stronger than vertical winds and horizontal scales are longer than vertical scales the turbulent shear production is actually further approximated by,
\begin{equation}
    S \approx \frac{\eta_t}{\rho_0} \left|\partial_z \uva\right|^2 \label{eq_shearprod_param}
\end{equation}

\subsubsection{Interaction between Parameterized Gravity Waves and Parameterized Turbulence}

For the analysis of the bidirectional interaction between turbulence and gravity waves through gravity wave induced turbulence production via shear and turbulent damping of gravity waves, take the system (\ref{eq_mean_mom}) -- (\ref{eq_shearprod_param}) as given, i.e. accept the approach of turbulence parameterization there, and split the mean flow into a mesoscale gravity-wave part and a synoptic-scale resolved-flow part. For ease of notation the angle brackets as indicators of the mean flow is dropped and re-introduce as indicating the synoptic-scale part, i.e., replace and decompose,
\begin{eqnarray}
    \vva &\rightarrow& \vv = \vva + \vv' \label{eq_decomp_v}\\
    \bta &\rightarrow& \bt = \bta + b'\\
    \Pta &\rightarrow& \Pt = \Pta + P'
\end{eqnarray}
Here the primes indicate the mesoscale fields. In principle, such a decomposition could also be made for the turbulent fields $K_t, \eta_t$, and $\kappa_t$. However, in the present study, the mesoscale fluctuations of turbulent fields are ignored and focus is instead shifted to the interaction between the synoptic-scale distribution of turbulence and mesoscale gravity waves.

For the discussion, \citeA{achatz_multi-scale_2023} is followed by first considering a locally monochromatic gravity-wave field and generalizing to a full spectral distribution afterwards, i.e., first assume,
\begin{equation}
    \left(\vv',b',P'\right) = \Re \left[\left(\vvh,\bh,\Ph\right) e^{i\Phi}\right] \label{eq_wkb}
\end{equation}
where the hatted quantities are complex amplitudes, and $\Phi$ is a phase defining a local wave number $\kv = \nabla\Phi$ and frequency $\omega = -\partial_t\Phi$. All amplitudes, the wave number, and the frequency have a synoptic-scale dependence on space and time. The mesoscale fields vanish on averaging in the phase over $2\pi$, e.g. by averaging over a wavelength or over a period, i.e. the synoptic-scale flow is the phase-averaged part of the flow and the angle brackets indicate a phase average.

What follows is an asymptotic analysis that is very close to the discussions of \citeA{achatz_multi-scale_2023}, so that here only the essential results are presented. Assume that viscosity and diffusivity, molecular and turbulent, are weak enough to not modify the gravity-wave dispersion relation,
\begin{equation}
    \omega = \kv_h\cdot\uva \pm \sqrt{\frac{N^2 k_h^2 + f^2 m^2}{k_h^2 + m^2}}  \label{eq_disp}
\end{equation}
and polarization relations,
\begin{eqnarray}
    \uvh &=& \frac{\kv_h\oh - i f \ez\times\kv_h}{\oh^2 - f^2} \left(1 - \frac{\oh^2}{N^2}\right) \frac{\bh}{i m} \label{eq_pol_u}\\
    \wh &=& \frac{i \oh}{N^2} \bh \label{eq_pol_w}\\
    \Ph &=& \left(1 - \frac{\oh^2}{N^2}\right) \frac{\bh}{i m}
\end{eqnarray}
where $\kv_h$ is the horizontal wave number, $k_h$ its absolute magnitude, $m$ the vertical wave number, and $\oh = \omega - \kv_h\cdot\uva$ the intrinsic frequency. From the dispersion relation (\ref{eq_disp}) also follow the eikonal equations,
\begin{eqnarray}
    \left(\partial_t + \cg\cdot\nabla\right) \kv 
    &=& 
    - \nabla\uva\cdot\kv \mp\ez d_z N^2 \frac{k_h^2/(2\sqrt{k_h^2 + m^2})}{N^2 k_h^2 + f^2 m^2} 
    \equiv \dot{\kv}
    \label{eq_eik_k}\\
    \left(\partial_t + \cg\cdot\nabla\right) \omega 
    &=& 
    \kv\cdot\partial_t\uva \label{eq_eik_o}
\end{eqnarray}

Inserting the wind decomposition (\ref{eq_decomp_v}) into the approximated shear-production rate (\ref{eq_shearprod_param}), using the WKB ansatz (\ref{eq_wkb}), and averaging over the phase, 
we obtain
\begin{eqnarray}
    S &=& \frac{\eta_t}{\rho_0} \left|\partial_z \uva\right|^2 + S_{gw} \\
    S_{gw} &=& \frac{\eta_t}{\rho_0} m^2 \frac{\left|\uvh\right|^2}{2}\label{eq: shear phase average}
\end{eqnarray}
Taking the polarization relations (\ref{eq_pol_u}) and (\ref{eq_pol_w}), starting with equation (\ref{eq_eik_k}) and using,
\begin{eqnarray}
\hat{b}^2=2\frac{N^4}{\bar{\rho}\hat{\omega}}\left(\frac{k_h^2}{k_h^2+m^2}\right)\mathcal{A}    
\end{eqnarray}
one gets,
\begin{eqnarray}
\bar{\rho}\frac{|\hat{u}|^2}{2}=\bar{\rho}\frac{\hat{u}\hat{u}^*}{2}=\frac{m^2}{\hat{\omega}k^2_h}\left(\frac{f^2l^2+k^2\hat{\omega}^2}{k^2_h+m^2}\right)\mathcal{A}\\
\bar{\rho}\frac{|\hat{v}|^2}{2}=\bar{\rho}\frac{\hat{v}\hat{v}^*}{2}=\frac{m^2}{\hat{\omega}k^2_h}\left(\frac{f^2k^2+l^2\hat{\omega}^2}{k^2_h+m^2}\right)\mathcal{A}\\ 
\end{eqnarray}
such that,
\begin{eqnarray}
 \bar{\rho}\frac{|\hat{\mathbf{u}}|^2}{2}=\bar{\rho}\left(\frac{|\hat{u}|^2}{2}+\frac{|\hat{v}|^2}{2}\right)=\frac{m^2}{\hat{\omega}}\left(\frac{f^2+\hat{\omega}^2}{k^2_h+m^2}\right)\mathcal{A} 
 \label{eq_ekin_gw}
\end{eqnarray}
where $\mathcal{A} = \mathcal{E}/\oh$ is the gravity-wave wave-action density and,
\begin{equation}
    \mathcal{E} 
    = \frac{\rho_0}{2} \left(\frac{\left|\vvh\right|^2}{2} + \frac{\left|\bh\right|^2}{2 N^2}\right)
\end{equation}
the gravity-wave energy density. A direct substitution of the horizontal-flow part (\ref{eq_ekin_gw}) of the GW kinetic energy into in equation (\ref{eq: shear phase average}) yields
\begin{equation}
    S_{gw} 
    = 
    \frac{2 \eta_t}{\rho_0^2} \frac{m^4 \left(f^2 + \oh^2\right)}{\oh \left(k_h^2 + m^2\right)} \mathcal{A}
    \label{eq_sgw}
\end{equation}
The time derivative of wave-action density is obtained by first deriving, in a manner parallel to the explanations by \citeA{achatz_atmospheric_2022} but respectively supplementing the momentum and entropy equations by turbulent friction and diffusion \cite<see also>{boloni_toward_2021}, a prognostic equation,
\begin{equation}
    \partial_t\mathcal{E} + \nabla\cdot\left(\cg\mathcal{E}\right) 
    + \kv_h\mathcal{A} \left(\cg-\uva\right) \cdot\cdot \nabla\uva
    =
    - 2 \left(k_h^2 + m^2\right) \frac{\rho_0}{2}
    \left(\eta_t\frac{\left|\vvh\right|^2}{2} + \kappa_t\frac{\left|\bh\right|^2}{2 N^2}\right)
\end{equation}
for the energy density. Together with the polarization relations (\ref{eq_pol_u}) and (\ref{eq_pol_w}) and the eikonal equations (\ref{eq_eik_k}) and (\ref{eq_eik_o}) this leads to,
\begin{equation}
    \partial_t\mathcal{A} + \nabla\cdot\left(\cg\mathcal{A}\right) 
    =
    - 2 \gamma \mathcal{A}
    \qquad
    \gamma = \left(k_h^2 + m^2\right) \left[\left(1 - \alpha\right)\eta_t + \alpha\kappa_t\right]
    \label{eq_wad}
\end{equation}
where,
\begin{equation}
    \alpha = \frac{N^2 k_h^2}{2 \oh^2 \left(k_h^2 + m^2\right)} \label{eq_alpha}
\end{equation}
is the fraction between gravity-wave available potential energy and total energy.

The gravity-wave shear production rate $S_{gw}$ in (\ref{eq_sgw}) describes the impact of the gravity waves on the synoptic-scale turbulence distribution, while the damping $\gamma$ in (\ref{eq_wad}) quantifies the impact of turbulence on the gravity-wave amplitudes. For final applicability in ICON/MS-GWaM, they are generalized in two steps. First, following \citeA{boloni_toward_2021}, incorporating the effects of density stratification on dispersion relation (\ref{eq_disp}), polarization relations (\ref{eq_pol_u}) and (\ref{eq_pol_w}), eikonal equation (\ref{eq_eik_k}), $S_{gw}$ in (\ref{eq_sgw}), and $\alpha$ in (\ref{eq_wad}) by replacing there $m^2$ by $m^2+\Gamma^2$ where $\Gamma^2$ is an anelastic-pseudo-incompressible modification of the squared vertical wave number, to be used in all instances named. This means especially that (\ref{eq_sgw}) and (\ref{eq_alpha}) are replaced by,
\begin{eqnarray}
    S_{gw} 
    &=&
    \frac{\eta_t}{\rho_0} m^2 \frac{2}{\rho_0} \frac{\left(m^2 + \Gamma^2\right) \left(f^2 + \oh^2\right)}{\oh \left[k_h^2 + \left(m^2 + \Gamma^2\right)\right]} \mathcal{A}
    \label{eq_sgw_an}\\
    \alpha 
    &=& 
    \frac{N^2 k_h^2}{2 \oh^2 \left[k_h^2 + \left(m^2 + \Gamma^2\right)\right]} 
    \label{eq_alpha_an}
\end{eqnarray}
Finally, allowance is also made possible for a continuous spectrum of gravity waves by introducing the spectral wave-action density $\mathcal{N}(\xv,\kv,t)$ so that $\mathcal{A} = \int dk^3\, \mathcal{N}$. As a result of this, the gravity-wave shear-production rate becomes,
\begin{equation}
    S_{gw} = \frac{\eta_t}{\rho_0} \int dk^3 m^2 \frac{2}{\rho_0} \frac{\left(m^2 + \Gamma^2\right) \left(f^2 + \oh^2\right)}{\oh \left[k_h^2 + \left(m^2 + \Gamma^2\right)\right]} \mathcal{N}
\end{equation}
and the prognostic equation for the wave amplitudes becomes,
\begin{equation}
\label{eq_gamma_final}
    \partial_t\Nc + \nabla_\xv\cdot\left(\cg\Nc\right) + \nabla_\kv\cdot\left(\dot{\kv}\Nc\right)
    =
    \partial_t\Nc + \cg\cdot\nabla_\xv\Nc + \dot{\kv}\cdot\nabla_\kv\Nc
    =
    - 2 \gamma \Nc
\end{equation}
Note that this $\gamma$ above (as defined in \ref{eq_wad}) is not local in spectral space, i.e., a GW will experience damping due to all the turbulence in its physical space, even the ones caused by other GWs. Furthermore, also note that for computational simplicity, in simulation runs at resolution used for this study, $\rho_0,N^2$ has been replaced by their resolved large-scale counterparts.
\subsection{Effects of Oblique Propagation}

As stated earlier, due to recent efforts from \cite{boloni_toward_2021,kim_toward_2021,kim_crucial_2024,voelker_ms-gwam_2024}, it has only just become possible in state-of-the-art gravity wave (GW) parameterization MS-GWaM to account for oblique propagation. However, conventionally, it is most common to assume a single-column path for the GWs, i.e., a 1D vertical propagation instead of a more physically accurate 3D 
propagation, and to neglect in the prognostic GW equations any horizontal gradients of the large scale flow.
This assumption then implies,
\begin{eqnarray}
    \label{eq_1d3d_wave_number_and_phase_speed}
    \dot{k}=\dot{l}=
    c_{g\lambda}=c_{g\phi}
    =0
\end{eqnarray}
where $k,l$ are the zonal and meridional wavenumbers, and 
$\mathbf{c}_{g\lambda},\mathbf{c}_{g\phi}$ 
are the zonal and meridional group velocity components respectively. Also set to 0 by definition are all horizontal derivatives
of GW wave-action density and all horizontal GW fluxes.
This then further reduces the wave action density budget of (\ref{eq_gamma_final}) to,
\begin{eqnarray}
    \label{eq_1d_gamma}
    \partial_t\Nc + \mathbf{c}_{gz}\partial_z\Nc + \dot{m}\partial_m\Nc
    =
    - 2 \gamma \Nc
\end{eqnarray}
Thereafter, this also reduces the contribution of GWs to the synoptic winds $\langle\mathbf{u}\rangle$ and potential temperature $\langle\theta\rangle$ via GW fluxes (entropy $\langle\mathbf{u}'\theta'\rangle$, and momentum $\langle\mathbf{u'v'}\rangle$) from,
\begin{equation}
\begin{split}
    \label{eq_3d_mean_mom}
    \left(\partial_t+\langle\mathbf{v}\rangle\cdot\nabla\right)\langle \theta\rangle&=-\nabla\cdot\langle\mathbf{u}'\theta'\rangle
    \\
    \left(\partial_t+\langle\mathbf{v}\rangle\cdot\nabla\right)\langle \mathbf{u}\rangle+f\mathbf{e}_z\times\langle\mathbf{u}\rangle&=-c_p\langle\theta\rangle\nabla_h\langle\pi\rangle-\frac{1}{\overline{\rho}}\nabla\cdot\left(\overline{\rho}\langle\mathbf{v'u'}\rangle\right)+\frac{f}{\overline{\theta}}\mathbf{e}_Z\times
    \langle\mathbf{u}'\theta'\rangle
\end{split}
\end{equation}
to,
\begin{equation}
\begin{split}
    \label{eq_1d_mean_mom}
    \left(\partial_t+\langle\mathbf{v}\rangle\cdot\nabla\right)\langle \theta\rangle&=0
    \\
    \left(\partial_t+\langle\mathbf{v}\rangle\cdot\nabla\right)\langle \mathbf{u}\rangle+f\mathbf{e}_z\times\langle\mathbf{u}\rangle&=-c_p\langle\theta\rangle\nabla_h\langle\pi\rangle-\frac{1}{\overline{\rho}}\partial_z\left(\overline{\rho}\langle w'\mathbf{u'}\rangle\right)
\end{split}
\end{equation}
where $\pi$ is the Exner pressure.

\subsection{Pseudomomentum Flux Approximation}
\label{sec:pm}

When the synoptic-scale flow is to leading order in geostrophic and hydrostatic equilibrium, it is completely defined to the same order by its quasigeostrophic potential vorticity $\zeta$, that satisfies the prognostic equation \cite<e.g.>{achatz_atmospheric_2022}
\begin{eqnarray}
    \label{eq_qgpv_pm}
    \left(\partial_t+\langle\mathbf{u}\rangle\cdot\nabla\right)\zeta=-\partial_x\left(\frac{1}{\overline{\rho}}\nabla\cdot\mathbf{\mathcal{H}}\right)+\partial_y\left(\frac{1}{\overline{\rho}}\nabla\cdot\mathbf{\mathcal{G}}\right)
\end{eqnarray}
with $\mathbf{\mathcal{G,H}}$ 
being the fluxes 
\begin{eqnarray}
    \mathbf{\mathcal{G}}
    &=& 
    \int d^3k\, k \Nc \hat{\mathbf{c}}_{g}\\
    \mathbf{\mathcal{H}}
    &=&
    \int d^3k\, l \Nc \hat{\mathbf{c}}_{g}
\end{eqnarray}
of the zonal and meridional components of pseudomomentum (PM),
$\mathbf{P}_h=\int d^3k\, \mathbf{k}_h\mathcal{N}$.
Notably, one can also arrive at the prognostic equation 
(\ref{eq_qgpv_pm}) 
with exactly the same GW forcing terms
when, instead of the direct GW fluxes of (\ref{eq_3d_mean_mom}), the PM 
fluxes are used in the momentum equation and the GW fluxes in the entropy equation are ignored, i.e.,
\begin{equation}
    \begin{split}
        \label{eq_3d_mean_mom_pm}
    \left(\partial_t+\langle\mathbf{v}\rangle\cdot\nabla\right)\langle \theta\rangle&=0
    \\
    \left(\partial_t+\langle\mathbf{v}\rangle\cdot\nabla\right)\langle \mathbf{u}\rangle+f\mathbf{e}_z\times\langle\mathbf{u}\rangle&=-c_p\langle\theta\rangle\nabla_h\langle\pi\rangle-\frac{1}{\overline{\rho}}\nabla\cdot\left(\mathbf{\mathcal{G}e}_x+\mathbf{\mathcal{H}e}_y\right)
    \end{split}
\end{equation}
\cite<e.g.>{wei_efficient_2019}.
This justifies the conventional ``pseudomomentum approach'' (\ref{eq_3d_mean_mom_pm}) for balanced flows.
While \citeA{wei_efficient_2019} found considerable differences between applications of the direct and the PM approach for imbalanced flow cases, the differences have been found to be smaller on average in global simulations using UA-ICON with $\sim$1$^\circ$ resolution \cite{voelker_ms-gwam_2024}.
Hence we also use the PM approach for simplicity. Additionally, under the pseudo-momentum approximation, the GW drag contribution in \ref{eq_gw_wave_drag} should be replaced by,
\begin{equation}
    \overline{\rho D_{gw}}
    =
    - \overline{\nabla\cdot\mathcal{G}}
\end{equation}

\subsection{ICON/MS-GWaM}

Building on \citeA{boloni_toward_2021,kim_toward_2021,voelker_ms-gwam_2024} and \citeA{kim_crucial_2024}, who introduced first the one dimensional gravity wave parameterization MS-GWaM-1D and then eventually its three dimensional extension MS-GWaM-3D,
the ICON model with its upper-atmosphere extension \cite{zangl_icon_2014,borchert_upper-atmosphere_2019} is chosen for the present work. Used is the version \textit{2.6.5-nwp1b} with a horizontal resolution of approximately 160~km (model grid labeled R2B04) and the physics package for numerical weather prediction (NWP) and the upper atmosphere. Note, that MS-GWaM replaces the non-orographic gravity wave drag but leaves the operational orographic wave parameterization intact \cite{lott_new_1997}
Note that with the current resolution, waves with horizontal wavelengths smaller than $\sim$1000~km are unresolved or significantly dissipated. Given this, we simply refer to the resolved waves as Rossby waves in our analysis.

The setup has a model top at an altitude of 150~km with vertical grid spacing of a few tens of meters in the boundary layer, 700-1500~m in the stratosphere, and a maximum of approximately 4~km in the lower thermosphere. A sponge layer acts above an altitude of 110~km which is why all analyses are restricted to altitudes below 100~km. The model is initialized with IFS analysis data below 60~km and with the climatological thermodynamic state at rest above. It is then spun up for a month to exclude adjustment effects in our analysis. 

\section{Experimentation}\label{sec:experiment}

This section is divided into two parts. Firstly, the impacts of oblique gravity wave propagation 
on the residual-mean circulation
are analyzed.
Afterwards, we couple the gravity wave parameterization to the turbulence scheme,
and its impact is further 
explored, with regard to both the mean circulation and turbulent mixing.
For both of them, ensemble sets for January over eight years are used,
initialized using the IFS analyses on 1st November of the years 1991--1998. Furthermore, the ensemble analyses uses a T-value of 1.76 corresponding to a $95\%$ confidence bound calculated similar to \citeA{Kyun_2015}.

\subsection{Oblique Gravity Waves Propagation}\label{exp:1d3d}

\citeA{voelker_ms-gwam_2024} have shown that oblique GW propagation has a strong effect on the global distribution of the GW fluxes and on the resulting GW drag. They demonstrate that this also leads to modified zonal-mean zonal winds and temperatures. In a companion study we will quantify the statistical significance of such differences, and investigate the consequences for solar tides. The present study supplements these investigations by addressing the residual mean circulation, GW shear and TKE.
As we adopt the pseudomomentum approach here (Section~\ref{sec:pm}) whereas the direct forcing approach was used in those previous studies,
we also show zonal mean winds and temperatures for completeness and better reference.
The results do not differ very much from those using the direct approach (not shown), so that the different approaches to GW forcing do not seem to contribute to leading order.

\subsubsection{Residual-Mean Circulation}

\begin{figure}
    \centering
    \includegraphics[width=0.8\textwidth]{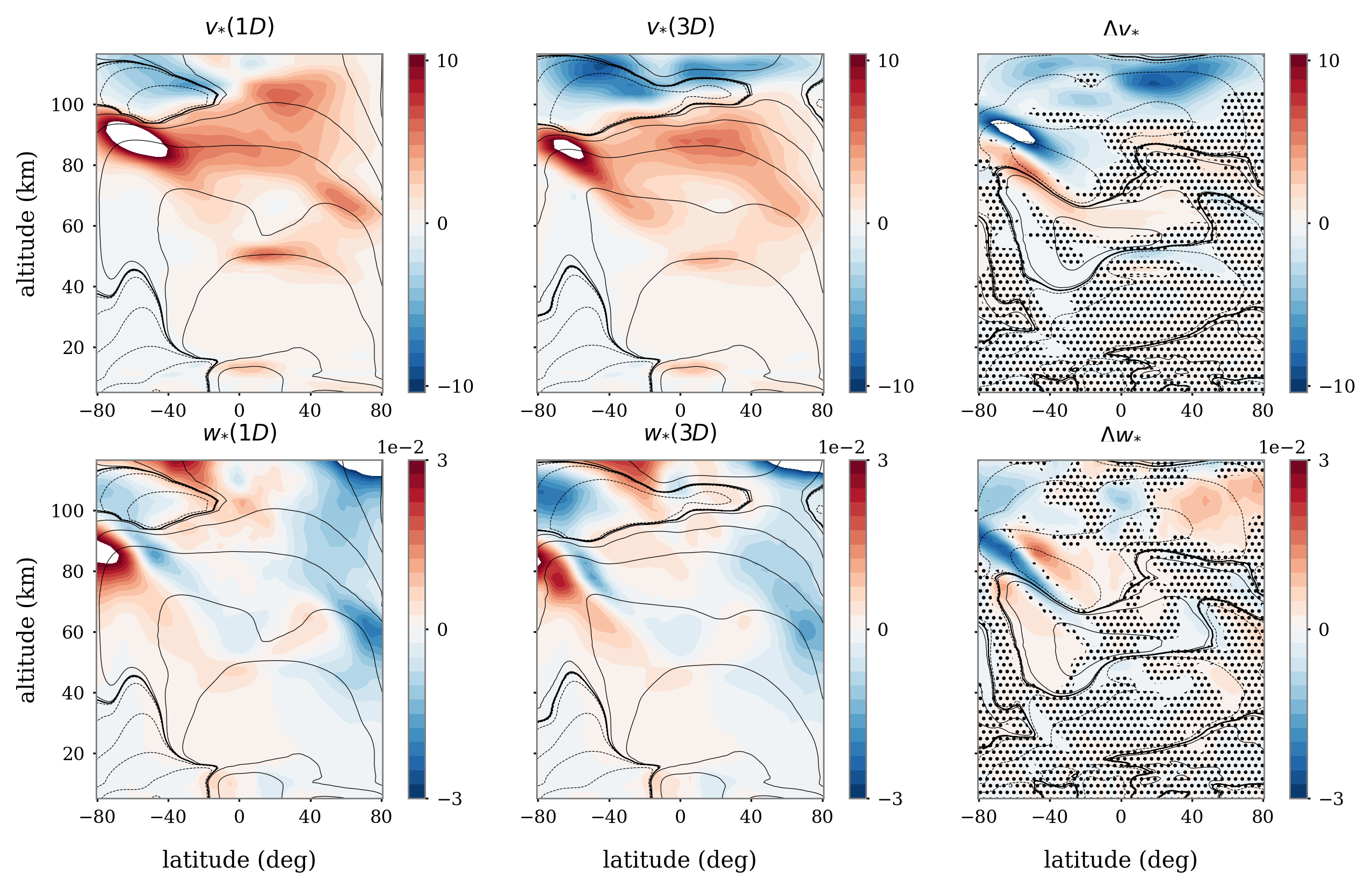}
    \caption{Plotted above is the zonal and monthly mean residual-mean circulation (RMC) for January of the ensemble set 1991 to 1998 using ICON-UA with MS-GWaM. Top row 
    shows
    the meridional 
    residual-mean wind
    ($m/s$) while the bottom 
    shows the vertical residual-mean wind
    ($m/s$). 
    In both rows the contour lines indicate the residual-mean streamfunction.
    The label ``1D" indicates 
    application of MS-GWaM as single-column
    GW parameterization with pseudomomentum approach, i.e., only vertical propagation of GWs allowed, ``3D" the case where oblique propagation is allowed as well, and the $\Delta$ performs 3D-1D.
    Stippling indicates statistical insignificance with 95\% statistical certainty, using a t value of 1.76.
    }
    \label{fig:tem120km_1d3d}
\end{figure}
As explained in Section (\ref{RMC}) the residual mean circulation (RMC) allows for diagnosis of the zonal-mean tracer transport while also accounting for the Stokes drift due to Rossby waves (RW) and gravity waves (GW). Shown in Figure \ref{fig:tem120km_1d3d} 
is a comparison between the January-mean RMC obtained by using MS-GWaM either in single column mode (1D) or allowing for oblique GW propagation (3D). Differences are shaded if their statistical significance is below 95\%.
The most prominent differences are seen in the mid-latitude and polar summer mesopause. The  maxima in upwelling and equatorward meridional motion are shifted downward by about 10~km. At the same time they are also weakened, from $3.32\times 10^{-2}$~m/s to $2.98\times 10^{-2}$~m/s in the vertical wind, and from 12.94~m/s to 10.52~m/s in the meridional wind. This is accompanied by an intensified circulation between about 40km altitude and the mesopause, with increased poleward-tilted upwelling especially in the mid-latitude summer mesosphere. Above the mesopause as well we see a relative circulation cell with rising air masses above the winter mesopause, summer-to-winter transport in the thermospere, sinking air masses above the summer pole and a closure of the circulation in the MLT. Moreover, in the upper and middle stratosphere one observes a reduction of the summer-hemisphere upwelling and the winter-hemisphere downwelling.

\begin{figure}
    \centering
    \includegraphics[width=0.7\textwidth]{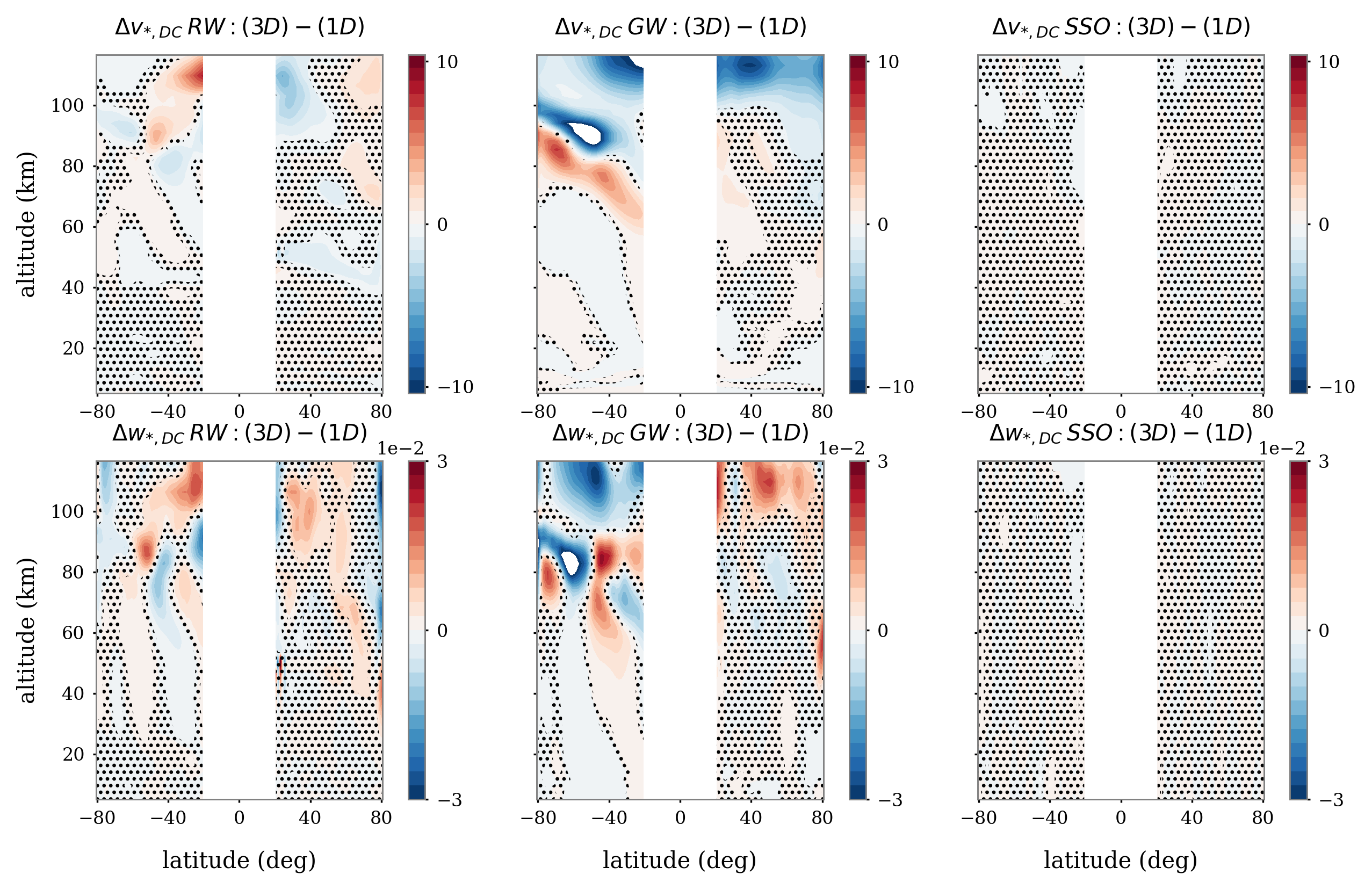}
    \caption{Plotted above 
    is the downward-control analysis of the impact of oblique GW propagation on the RMC in the
    the ensemble set 1991 to 1998 using ICON-UA with MS-GWaM. Top row 
    shows results for the residual-mean merdional wind, and the bottom row for the vertical wind.
    The label ``1D" indicates 
    application of MS-GWaM in single-column mode with
    pseudomomentum approach, i.e., only vertical propagation of GWs allowed, ``3D" the case where oblique propagation is allowed as well, and the $\Delta$ 
    indicates the difference between 3D and 1D.
    The left column shows the contribution to changes 
    by a re-distribution of Rossby waves (RW), the center column the contribution to changes in the non-orographic gravity waves (GW) simulated by MS-GWaM, and the right the contribution to changes in subgrid-scale-orographic gravity waves (SSO), following the parameterization of \citeA{lott_new_1997}.}
    \label{fig:tem120km_1d3d_decomp}
\end{figure}
Figure \ref{fig:tem120km_1d3d_decomp} gives a first interpretation of the observed differences using downward control. It shows the contribution to the changes in the RMC by resolved Rossby waves (RW), the non-orographic GWs simulated by MS-GWaM (GW), and also a negligible contribution from the parameterization of orographic GWs by the scheme of \citeA{lott_new_1997}. Ones sees that the modified forcing by non-orographic GWs is responsible for both the downward shift of the regions of strongest residual-mean winds at the mesopause and the enhanced poleward-tilted upwelling in the mid-latitude summer mesosphere below the mesopause. The re-distribution of the resolved Rossby waves tends to mitigate this effect, in line with corresponding findings of \citeA{cohenWhatDrivesBrewer2014}. The inverse relative circulation cell above the mesopause can also be attributed to the modified GW forcing, with also here the RW forcing mitigating the effect somewhat. Finally we also note that the reduction in the winter-hemisphere downwelling can be attributed to the changes in the GW forcing as well, now however without a corresponding mitigation by resolved eddies.

Certainly the analysis above does not yet explain why the RW and GW fluxes re-configure as they do. One indication with regard to the summer mesosphere and mesopause can perhaps be drawn from the findings of \citeA{voelker_ms-gwam_2024} that GWs tend to be focused not only into the polar night jet, but also into the easterly jet in the summer mesosphere, in agreement with expectations from the GW dispersion relation. This might contribute to enhancing the GW forcing below the mesopause and reducing it higher up.

\subsubsection{Zonal-Mean Temperature and Zonal Wind}

\begin{figure}
    \centering
    \includegraphics[width=0.9\textwidth]{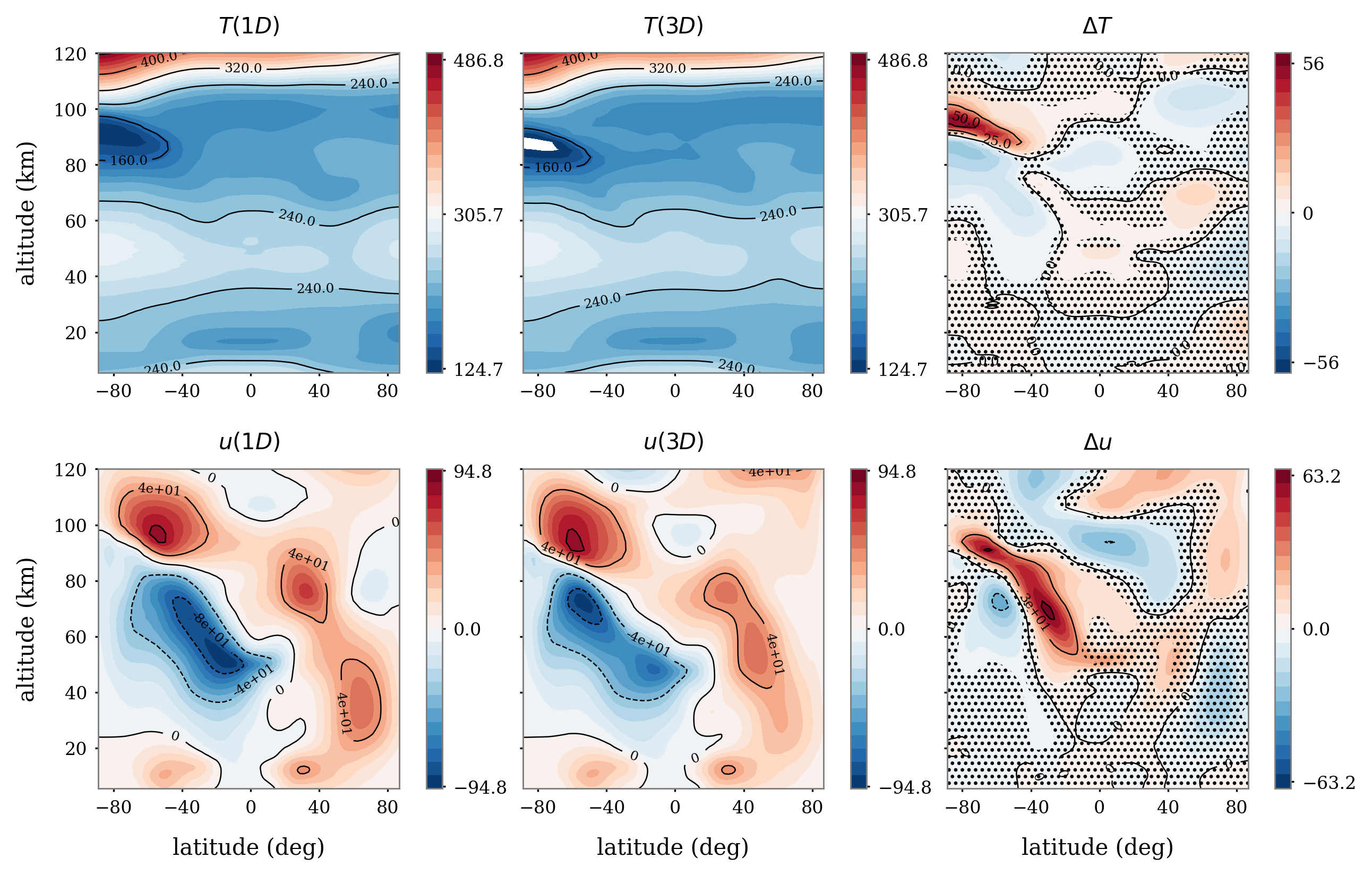}
    \caption{Plotted above are the zonal 
    and
    monthly mean profiles for January of the ensemble set 1991 to 1998 using ICON-UA with MS-GWaM. 
    Top row 
    shows
    the temperature ($^oK$) while the bottom 
    shows 
    the zonal wind ($m/s$). The label ``1D" indicates 
    application of MS-GWaM in single-column mode 
    with pseudomomentum approach, i.e., only vertical propagation of GWs allowed, 
    and
    ``3D" the case where oblique propagation is allowed as well, and the $\Delta$ 
    indicates the difference 3D-1D.
    }
    \label{fig:temp-u_120km_1d3d}
\end{figure}
Inspecting the 
differences in the
zonal 
and
monthly mean temperature 
in Figure \ref{fig:temp-u_120km_1d3d}, 
one observes as a consequence of oblique GW propagation
a downward shifting of the summer mesopause (down to around $85$km
over the south pole), in line with the downward shift in the maxima of residual-mean upward and equatorward circulation diagnosed above.
One also sees that the lowered polar summer mesopause is cooler. We attribute this to the enhanced upwelling in the mid-latitude summer mesosphere. Likewise the relative inverse circulation in the thermosphere leads to a relative cooling of the winter lowermost thermosphere, and a relative heating on the summer side.  Moreover, the weakened downwelling in the winter-hemisphere stratosphere seems to directly lead to a relative cooling in this region.

The zonal-mean zonal-wind plots
in Figure \ref{fig:temp-u_120km_1d3d} show an impact of oblique GW propagation that can mostly be inferred via the thermal-wind relation from the impact on the zonal-mean temperature field. The lowered thermal summer mesopause directly leads to a lower wind mesopause with strong shears and a wind-direction reversal. Likewise the cooled northern-hemisphere mid-latitude stratosphere goes along with an equatorward shift of the stratospheric polar-night jet.

\subsubsection{GW Shear and TKE}

\begin{figure}
    \centering
    \includegraphics[width=0.9\textwidth]{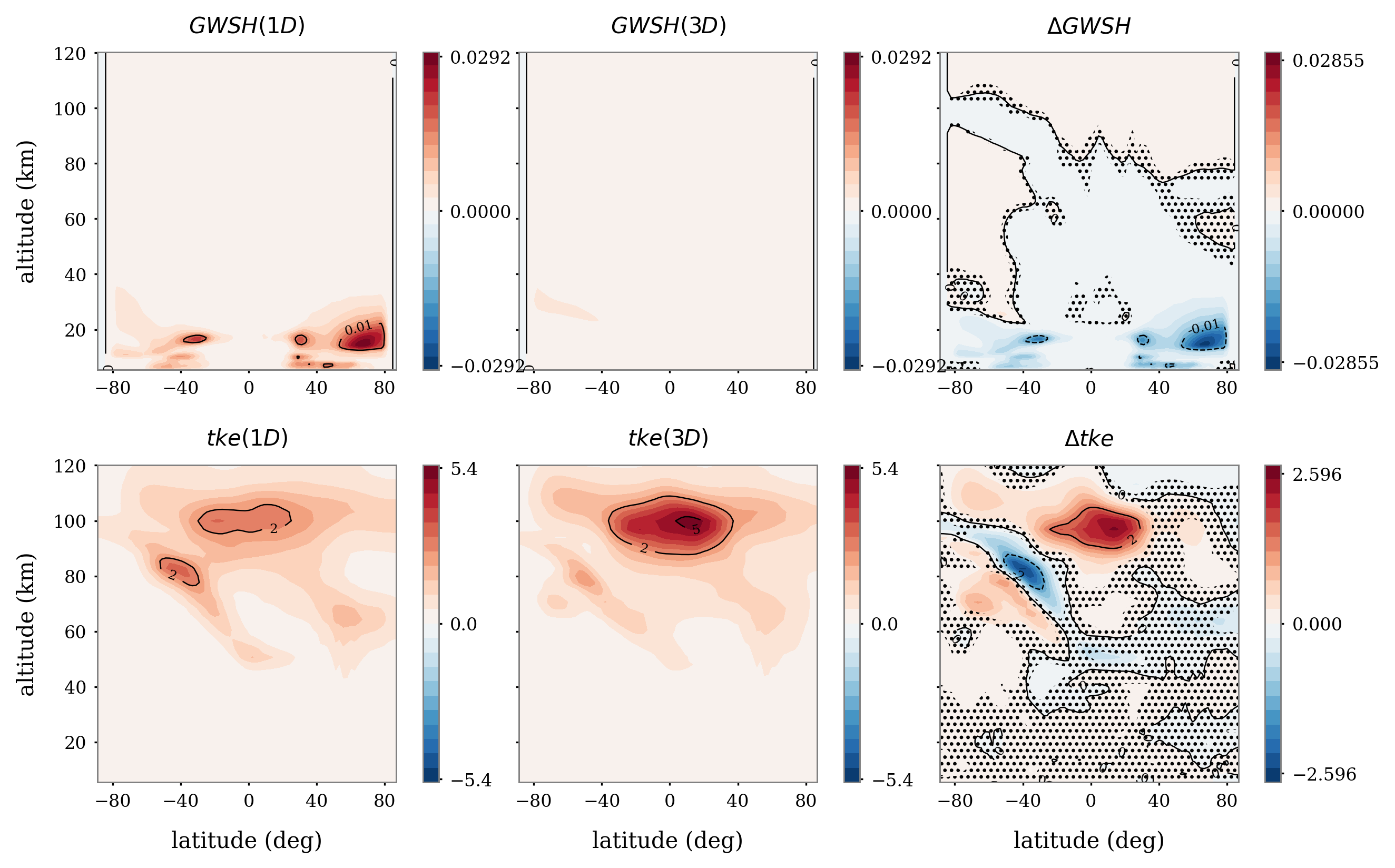}
    \caption{Plotted above are the zonal 
    and
    monthly mean profiles for January of the ensemble set 1991 to 1998 using ICON-UA with MS-GWaM. Top row 
    shows the shear due to parameterized
    GWs ($s^{-2}$) while the bottom 
    shows the turbulent kinetic energy ($m^2/s^2$). The label ``1D" indicates 
    application of MS-GWaM in single-column mode 
    with pseudomomentum approach, i.e., only vertical propagation of GWs allowed, ``3D" the case where oblique propagation is allowed as well, and the $\Delta$ 
    indicates the difference
    3D-1D.}
    \label{fig:shear_tke_120km_1d3d}
\end{figure}
Finally, looking at the shear induced by gravity waves $\langle|\partial_z\hat{\mathbf{u}}|^2\rangle$ and turbulent kinetic energy $|\mathbf{v}'|^2/2$ in Figure \ref{fig:shear_tke_120km_1d3d}, the oblique propagation of GWs is observed to reduce the shear, especially in the lower atmosphere and mid to high latitudes by at least 90\%. 
A cause for this might be that the horizontal propagation helps GWs avoiding critical layers where their vertical wavelength is reduced most effectively.
Close to the summer mesopause one sees a downward shift of the GW shear, fully in line with a downward shift of the mesopause wind reversal and the corresponding enhancement of the GW shear.
For TKE, again the changes observed are significant with equatorial lower thermosphere experiencing an enhanced turbulence by nearly 70\%.
Moreover, the localized region of 
the summer mesosphere 
experiences 
a downward shift of the TKE maximum that seems to go along with the downward shift of the zonal-mean zonal-wind reversal. All of these impacts are without direct coupling between parameterized turbulence and parameterized GWs.

\subsection{Coupling between Gravity Waves and Turbulence}

Moving to changes due to 
the interaction between GWs and turbulence,
this section examines the impact of bidirectionally coupling the GW parameterization MS-GWaM-3D and the turbulence scheme of 
UA-ICON
as per the developments of section \ref{sec: GWT coupling}.

\subsubsection{Residual-Mean Circulation}

\begin{figure}
    \centering
    \includegraphics[width=0.9\textwidth]{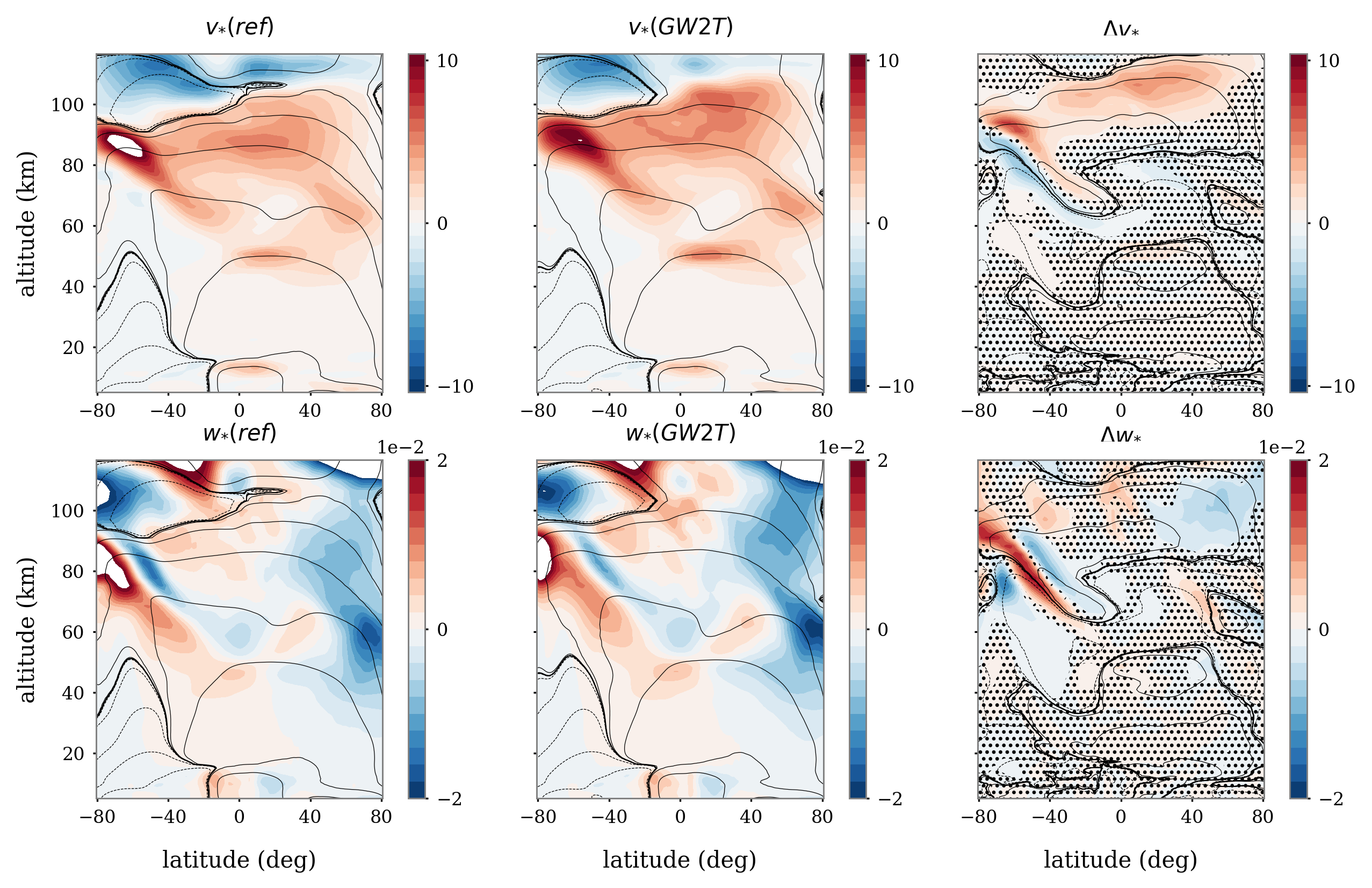}
    \caption{
    As Fig. \ref{fig:tem120km_1d3d}, but now showing the difference ($\Delta$, right column) in the residual-mean circulation between simulations without (left, ref) and with (middle, GW2T) bidirectional coupling between MS-GWaM-3D and the TKE turbulence parameterization.
    }
    \label{fig:tem120km_GW2T}
\end{figure}
Following the same analysis from Section \ref{exp:1d3d}, shown in Figure \ref{fig:tem120km_GW2T} 
is the impact of the coupling on the RMC. Most striking is that here, and also in the results to follow, the response in the mesosphere and thermosphere is almost opposite to the one from oblique GW propagation.
The regions of strongest upwelling and equatorward transport in the summer MLT are shifted upwards. Above the mesopause one observes relative circulation with upwelling over the summer pole, winterward transport in the thermosphere, downwelling above the winter pole and a closure by summerward transport in the MLT. Below the mesopause one obtains relative upward poleward-tilted motions. 
The effect on the summer mesopause is within expectation, as coupling the turbulence to the GWs makes the waves feel damping due to turbulence throughout their lifetime. This then allows them to propagate further up without 
breaking.

\begin{figure}
    \centering
    \includegraphics[width=0.9\textwidth]{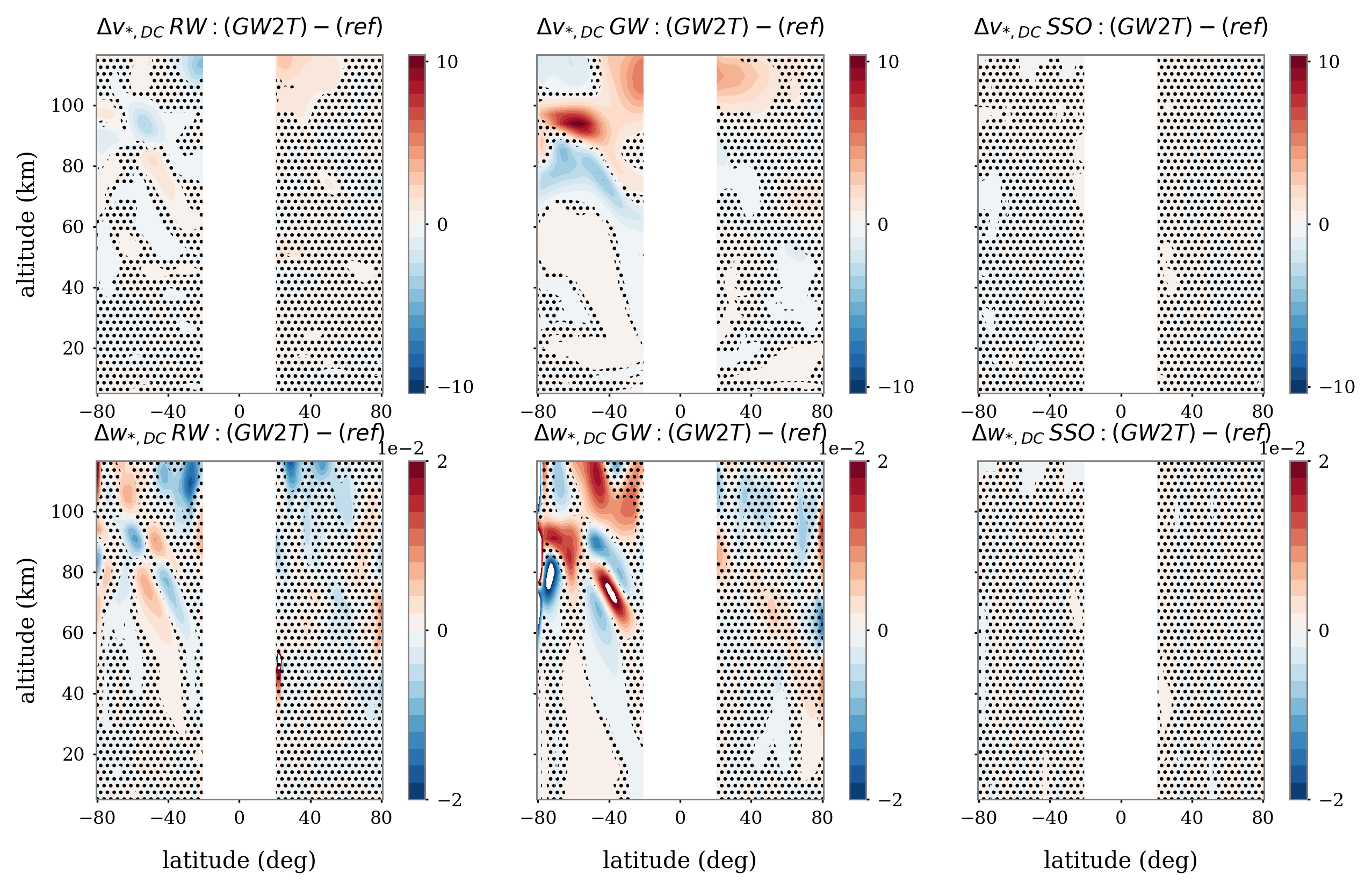}
    \caption{
    As Fig. \ref{fig:tem120km_1d3d_decomp}, but now analysing instead of the effect of oblique GW propagation the impact of the bidirectional coupling between MS-GWaM-3D and the TKE turbulence parameterization.
    }
    \label{fig:dc_120km_GW2T}
\end{figure}
Using downward control analysis again, shown in Fig. \ref{fig:dc_120km_GW2T}, and comparing to Fig. \ref{fig:tem120km_GW2T} it is evident that the circulation changes can be attributed predominantly to the modified GW forcing, with a weaker mitigating influence from the resolved Rossby waves. Again, mountain waves do not contribute in a relevant manner.

\subsubsection{Zonal-Mean Temperature and Zonal Wind}

\begin{figure}
    \centering
    \includegraphics[width=0.9\textwidth]{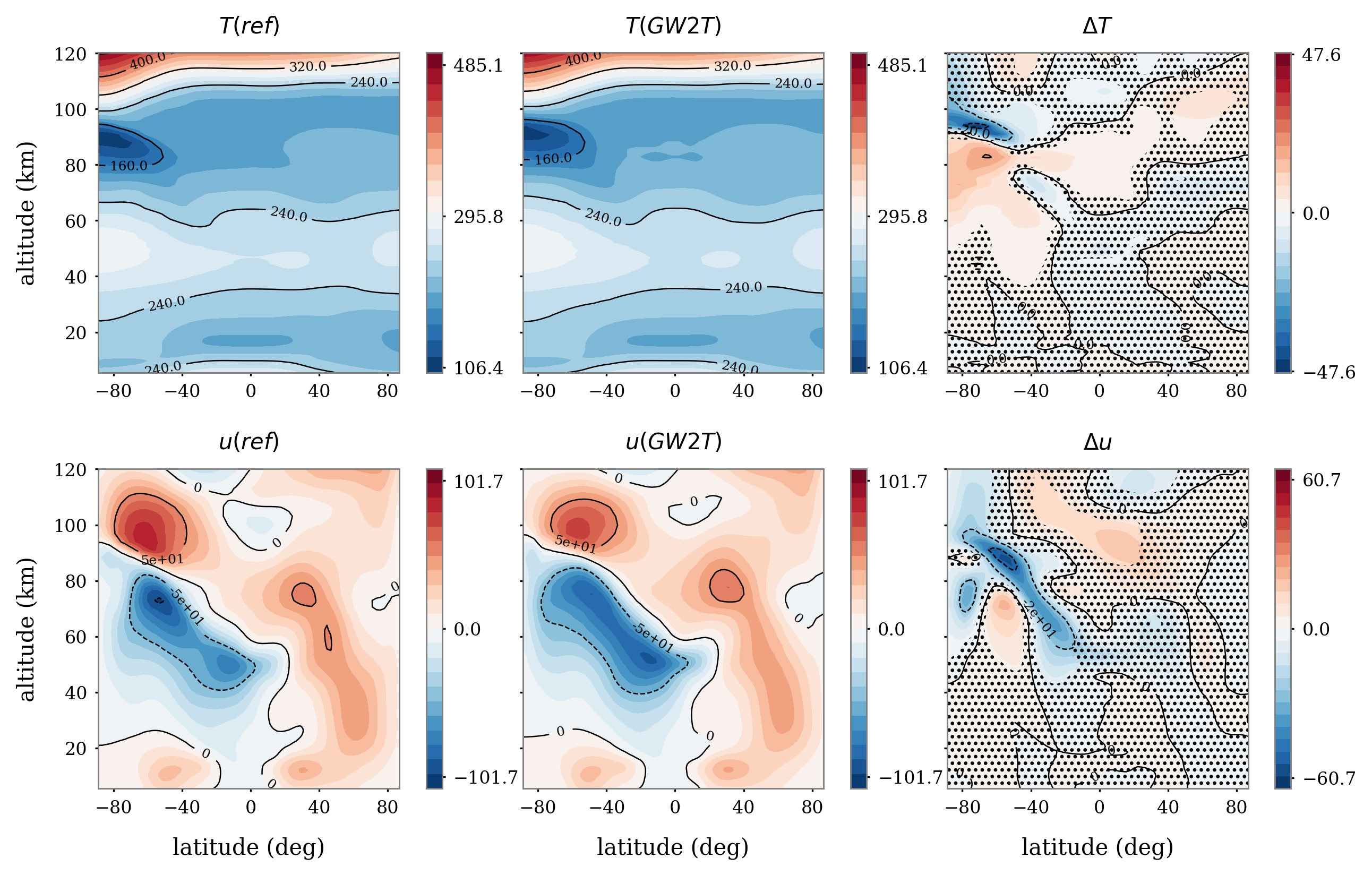}
    \caption{
    As Fig. \ref{fig:temp-u_120km_1d3d}, but now showing instead of the impact of oblique GW propagation the effect of the bidirectional coupling between MS-GWaM and the TKE turbulence parameterization.
    }
    \label{fig:temp-u_120km_GW2T}
\end{figure}
Inspecting the zonal 
and 
monthly mean temperatures in Figure \ref{fig:temp-u_120km_GW2T}, the summer mesopause indeed appears to be shifted back upwards to about $90$~km 
over the summer pole. As compared to the case without GW-turbulence coupling, where its temperature is around $107~^o$K, it is now also warmer again, with a temperature of around $112~^o$K.
This can be understood by the upward shift of the regions of strongest upwelling and equatorward transport and the weakened upwelling in the mid-latitude mesosphere below the mesopause.
Directly from this 
follows, via thermal-wind balance,
a downward shift of the altitude of mesopause zonal-wind reversal, and also a slight upward shift of the tropical wind reversal in the lower thermosphere. As compared to the effect of oblique GW propagation, however, the effect in the stratosphere seems to be weaker.

\subsubsection{GW Shear and TKE}

\begin{figure}
    \centering
    \includegraphics[width=0.9\textwidth]{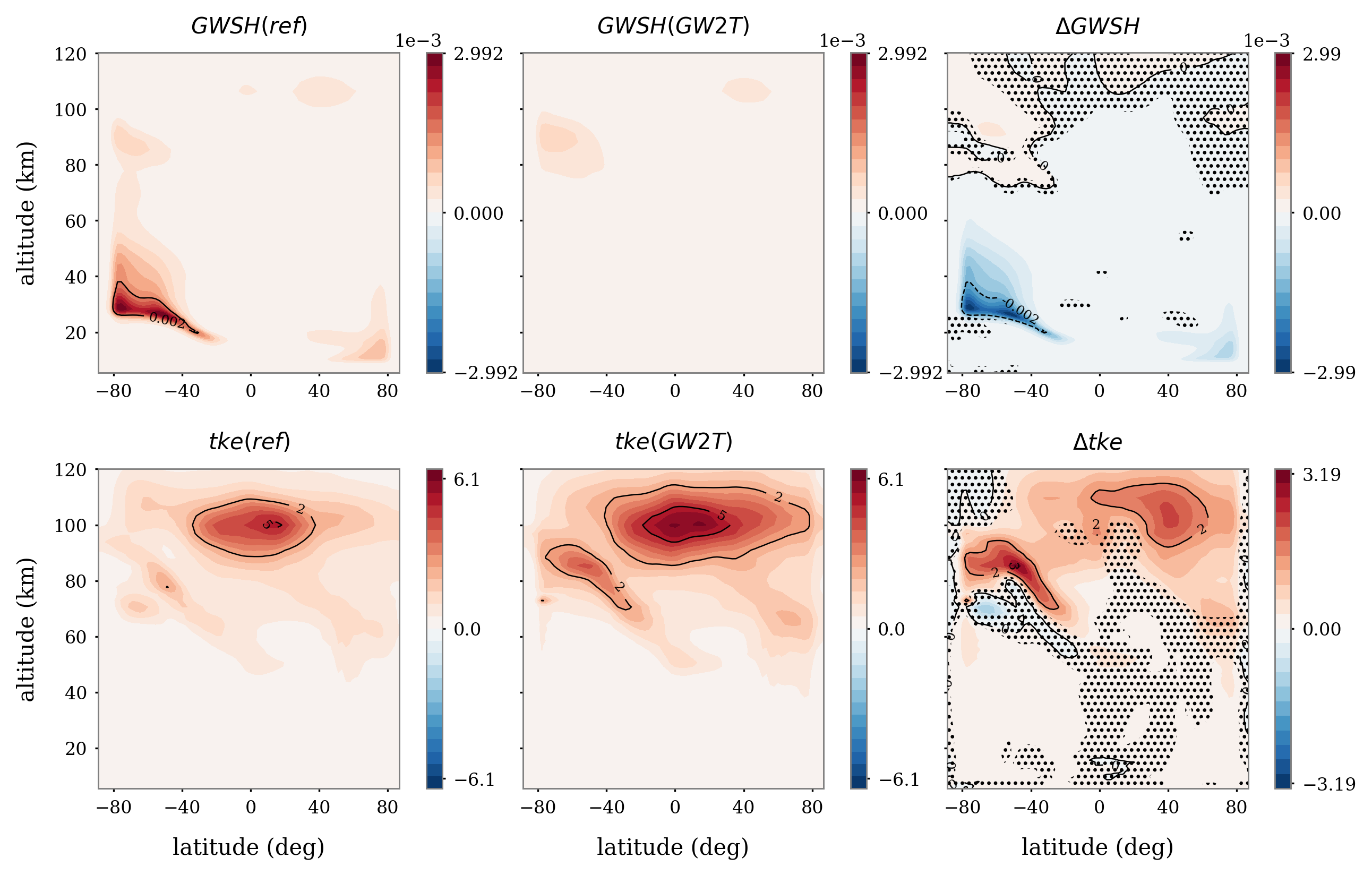}
    \caption{
    As Fig. \ref{fig:shear_tke_120km_1d3d}, but now showing instead of the impact of oblique GW propagation the effect the bidirectional coupling between MS-GWaM and the TKE turbulence parameterization.
    }
    \label{fig:shear_tke_120km_GW2T}
\end{figure}
Looking at the fields most directly impacted by the bidirectional coupling, i.e., the GW induced shear and TKE in Figure \ref{fig:shear_tke_120km_GW2T}, in this case it does not reverse the effects due to 3D propagation unlike in the case of RMC, temperature and zonal winds, but rather mirrors it further. Here too, the shear is almost globally reduced and almost completely (over 75\%) from the stratopause and below and in mid to high latitudes. 
The GW shear aligned with the mesopause wind reversal, however, is not reduced, but rather shifted upwards together with zonal-mean zonal wind shear.
Turbulence
is observed to be enhanced 
more or less throughout the atmosphere. Due to GW shear production at the summer mesopause, TKE is increased from about 2 to 5~m$^2$/s$^2$, and it is also shifted upwards, following the location of the summer-mesopause wind reversal. Moreover, a considerable TKE enhancement is also observed as well in the whole lower thermosphere as in the stratosphere.
These observations are in line with expectations as the coupling introduces an additional source of turbulence, while it also damps the GWs which then reduces the shear that they induce. Only in the summer-mesopause shear due to the large-scale wind reversal the turbulent reduction of the GW shear is balanced by an enhancement by the large-scale shear.

\subsubsection{Turbulent Tracer Mixing}

\begin{figure}
    \centering
    \includegraphics[width=0.9\textwidth]{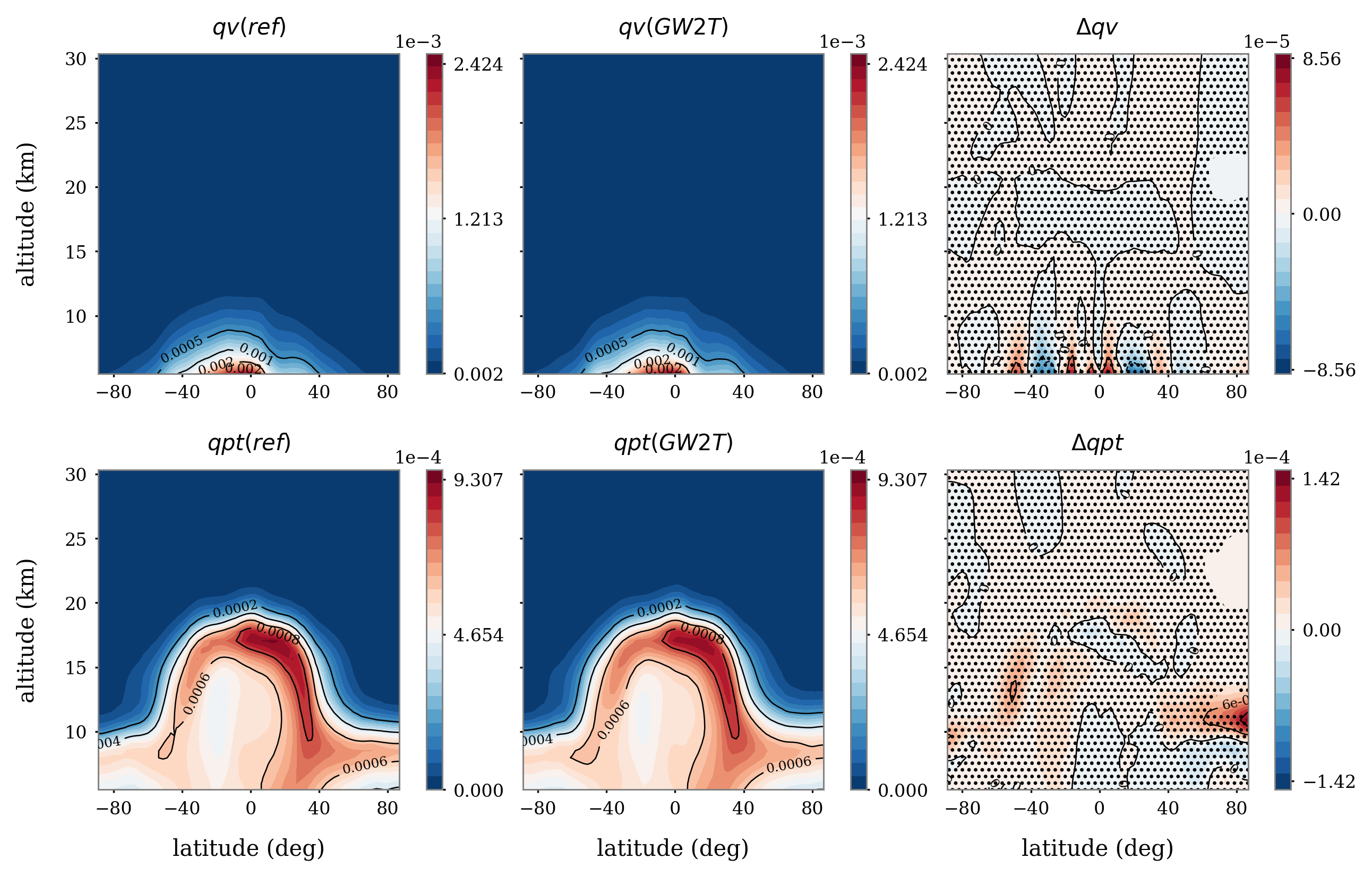}
    \caption{Plotted above are 
    zonal 
    and monthly mean tracer-mixing-ratio profiles for January of the ensemble set 1991 to 1998 using ICON-UA with MS-GWaM-3D. Top row 
    shows the 
    mixing ratio of water vapor while the bottom is that of 
    a passive tracer. The label ``ref" indicates 
    the result from simulations without coupling between MS-GWaM and TKE turbulence parameterization, while 
    ``GW2T" indicates 
    results from simulations with this coupling, 
    and the $\Delta$ 
    indicates the difference due to the bideirectional coupling.
    Note that for these simulations the passive tracer has been initialized on Nov 1st by the water-vapour 
    mixing ratio.
    }
    \label{fig:qv_qpt_30km_GW2T}
\end{figure}
Finally we have a look at the impact of the increase of TKE in the stratosphere on the transport and mixing of tracer substances. For this purpose we show the zonal-mean mixing ratios, again  in January, of water vapor and of a passive tracer that has been initialized on November 1st by the water vapor mixing ratio. As is to be seen in 
Figure \ref{fig:qv_qpt_30km_GW2T}, a statistically significant change 
can only be 
observed for the passive tracer,
with an increase of about 15\%
near the northern 
tropopause, in the altitude and latitude range of 
$10-15$~km and 
$55-80^o$N, respectively.
The lack of an impact on water vapor seems to indicate that freeze-drying at the tropopause dominates over any effect of cross-tropopause mixing. The effect on a passive tracer, however, is larger than for water vapor. Given the short model integration time here together with the largest effects regarding the impact of turbulence at higher altitudes, we speculate that there are larger effects on tracer distributions at higher altitudes. However, given the transport timescales from the lower into the upper stratosphere and lower mesosphere, this requires substantially longer model integrations than have been completed here. This will be analyzed further in subsequent studies.

\section{Summary and Discussion}\label{sec:summary}

In summary, this work explores the impact of gravity waves (GW), their oblique propagation, and their coupling to turbulence on transport and mixing. It provides a
step towards a
theoretical framework and demonstrates meaningful and statistically significant 
impacts within the upper atmosphere extension UA-ICON of the German community climate model ICON, using MS-GWaM for the parameterization of non-orographic GWs. 
Allowing for oblique propagation is 
shown, among other effects,
to lower and cool the summer mesopause, reduce the 
shear due to small-scale GWs
(in tropopsphere and most of the middle atmosphere)
and enhance the turbulence (in the upper atmosphere). 
Bidirectional coupling between parameterized GWs and turbulence is introduced, via a source of TKE by GW shear and turbulent damping of GWs. Among other effects this 
is shown to lift and warm the summer mesopause, but again reduce the GW induced shear 
(in the middle and lower atmosphere) 
and enhance the turbulence 
(almost in the whole atmosphere). 
It is also shown that the enhanced turbulent mixing has a significant effect on the zonal-mean mixing ratio in the UTLS of a passive tracer that has originally been initialized in the troposphere, with a mixing ratio identical to that of water vapor.

To conclude, 
both oblique GW propagation and the interaction between GWs and turbulence seem to have a relevant impact on the transport and mixing of tracer substances that ought to be respected in subgrid-scale parameterizations. 
However, it still remains to be seen how well 
our findings
compare to observational data. 
Potentially, the GW source spectra will have to be adjusted for optimal agreement, as well as critical parameters in the turbulence parameterization. In that regard we also note that the TKE turbulence parameterization itself is not the last word spoken, as an inclusion of turbulent available potential energy might be essential to close the energy budget \cite{bastak_duran_two-energies_2022}. 
Equally valid are questions on the scale awareness and stability of this new coupled 
system. It might be interesting, e.g., to increase the mathematical accuracy of the handling of GWs close to the poles, an issue already mentioned by \citeA{voelker_ms-gwam_2024}. Finally it also remains an open question what impact one obtains from orographic GWs once those have been integrated into MS-GWaM.
We admit the importance of these questions and will address them in a future work. This manuscript rather aims to establish 
a first
theoretical foundation and report on the corresponding observed 
effects
only.

\clearpage
\appendix
\section{Implementation within ICON-MSGWaM}\label{appendix: icon code}
In interest of reproducibility and wider adoption, it is useful to describe the implementation of the above discussed theory within the ICON-MSGWaM framework. The necessary changes can be broadly grouped into two parts.
\begin{figure}[h]
\centering
 \noindent\includegraphics[width=0.9\textwidth]{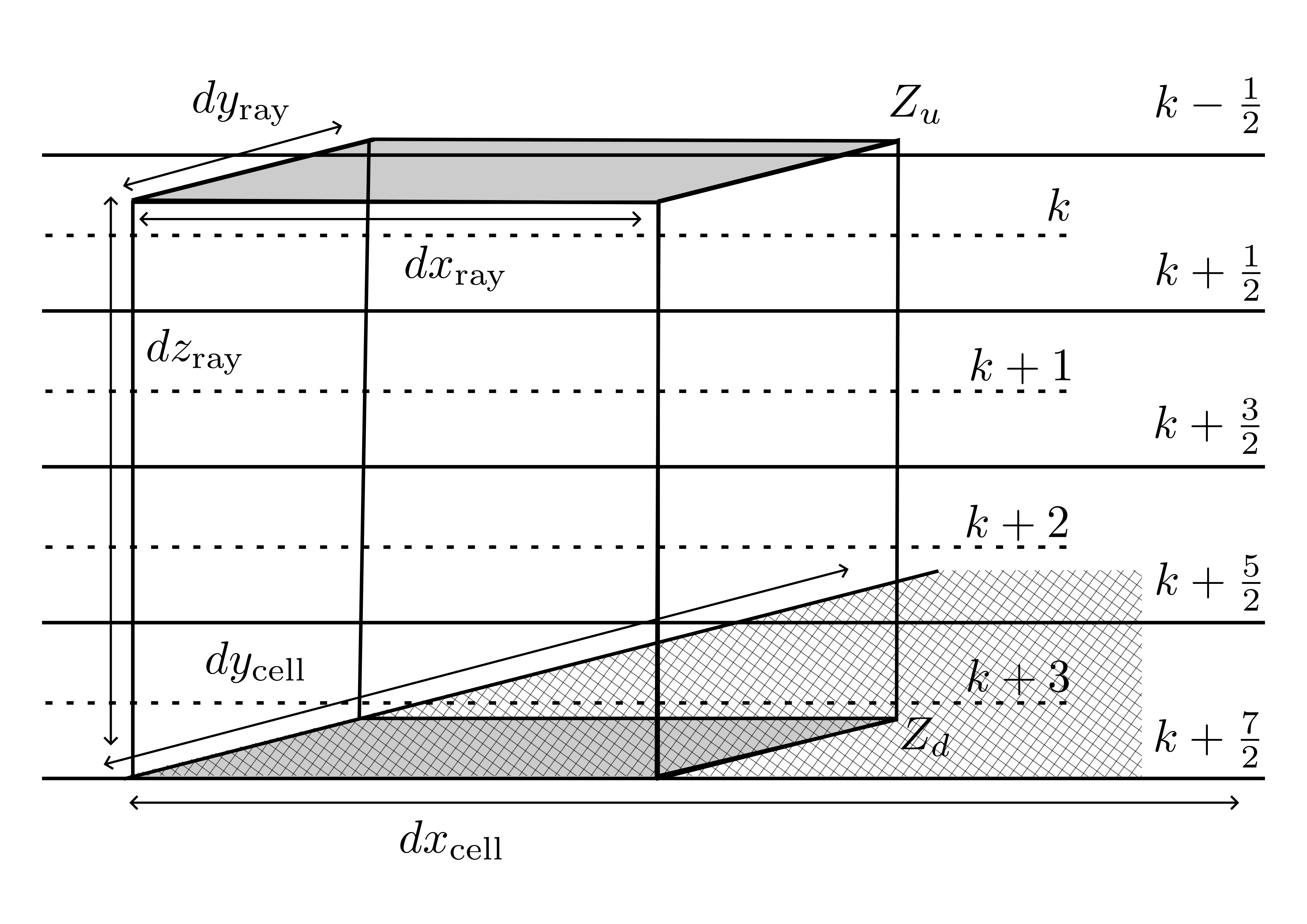}
 \vspace{-3mm}
\caption{Illustration of a ray volume in relation to vertical layers and cell area in ICON-MSGWaM setup. Layer interfaces are defined by half levels and layer centers by full levels. The full (or half) level at the top of the atmosphere is indexed by $k=1$ and it increases downwards. The top of the ray volume is indexed by $Z_u$ and the bottom with $Z_d$. Each ray volume can span a (non-integer) number of vertical layers.}
\label{fig_ray_vol}
\end{figure}
\subsection{Damping}
This part explains the portion of the code that deals with implementing the damping of gravity waves due to turbulence. For a variable ${X}$, it is first interpolated to the ray center as,
$$
X_{c,r}=X_{c,k+1/2}+\left(\frac{X_{c,k-1/2}-X_{c,k+1/2}}{z_{c,k-1/2}-z_{c,k+1/2}}\right)\left(z_{c,r}-z_{c,k+1/2}\right)
$$
where $z_{c,r}$ is the height at center of the ray volume, i.e., $z_{c,r}=[Z_d+(Z_u-Z_d)/2]_{c,r}$ as per Figure \ref{fig_ray_vol} and index $c$ represents cell location while indices $r,k+1/2$ represents the vertical location of either ray volume center or layer interface $k+1/2$ respectively. In case of ICON-MSGWaM in relation to the turbulence-gravity wave coupling, the necessary quantities being interpolated are,
$$
X=\left[\eta_t,\kappa_t,\gamma_m,N^2,f^2\right]^T
$$
where $\eta_t,\kappa_t$ are the turbulent viscosity and diffusivity respectively, $\gamma_m$ the damping due to molecular diffusivity, $N^2$ the buoyancy frequency, and $f^2$ the coriolis frequency. Then, the factor $\alpha$ at center of the ray volume in a given cell is computed as,
$$
\alpha_{c,r}=\frac{(N^2k^2_h)_{c,r}}{2\tilde{\hat{\omega}}^2_{c,r}(k^2_h+m^2+\Gamma^2)_{c,r}}\approx
\frac{(N^2k^2_h)_{c,r}}{2f^2_{c,r}(m^2+\Gamma^2)_{c,r}+(N^2k^2_h)_{c,r}}
$$
where the corrected intrinsic frequency $$\tilde{\hat{\omega}}^2_{c,r}\approx \left[f^2_{c,r}(m^2+\Gamma^2)_{c,r}+(N^2k^2_h)_{c,r}\right]/\left[(k^2_h+m^2+\Gamma^2)_{c,r}\right]$$ for gravity waves. With this, the damping $\gamma_t$ due to turbulent viscosity and diffusivity at center of the ray volume is then,
$$
{\gamma_t}_{c,r}=\left[\eta_t(1-\alpha)+\alpha\kappa_t\right]_{c,r}(k^2_h+m^2)_{c,r}
$$
Which is then added with the damping due to molecular viscosity such that $\gamma_{net}=\gamma_t+\gamma_m$. Finally, the damping of the wave action density (calculated at its center) is then implemented as a second-order in time solution to $d\mathcal{N}/dt=-2\gamma_{net}\mathcal{N}$ as,
$$
\mathcal{N}_{c,r}^{n+1}=\mathcal{N}^{n}_{c,r}e^{-2\gamma_{net}^{n+1/2}dt}=\mathcal{N}^{n}_{rc}e^{-(\gamma_{net}^{n+1}+\gamma_{net}^{n})_{c,r}dt}
$$
\subsection{Shear}
This part explains the portion of the code which implements the production of turbulence due to gravity waves through the shear production term. Similar to the damping part, here too the corrected intrinsic frequency for gravity is computed but at layer center $k$ as, 

$$\tilde{\hat{\omega}}^2_{c,k} (r) = \left[f^2_{c,k}(m^2_{c,r}+\Gamma^2_{c,k})+(N^2k^2_h)_{c,r}\right]/\left[({k^2_h}_{c,r}+m^2_{c,r}+\Gamma^2_{c,k})\right]$$

Note that here, $(r)$ shows that it is unique to a ray volume and will be different for others. Furthermore, the total wavenumber $K$ and vertical wavenumber $m$ is also amended with the anelastic correction at layer centers as, 
$$\tilde{m}^2_{c,k}(r)=m^2_{c,r}+\Gamma^2_{c,k}$$
$$\tilde{K}^2_{c,k}(r)=K^2_{c,r}+\Gamma^2_{c,k}$$
Next, the partial wave action density $A$ of a ray volume in a cell in relation to the vertical layer $k$ and covering the vertical extent $k-1/2$ to $k+1/2$ is calculated as,
$$
A_{c,k}(r)=\mathcal{N}_{c,r}\left(dkdldm\right)_{c,r}\frac{(dxdy)_{c,r}}{(dxdy)_{c,k}}\frac{\left(dz^{Z_u}_{Z_d}\right)_{c,r}\cap\left(dz^{k-1/2}_{k+1/2}\right)_{c,k}}{\left(dz^{k-1/2}_{k+1/2}\right)_{c,k}}
$$
here, the multiplication by $(dkdldm)$ collapses the phase space to physical space spanning the cell. Then, the multiplication by ${(dxdy)_{c,r}}/{(dxdy)_{c,k}}$ further collapses the physical space spanning the horizontal cell area to only spanning the horizontal area of the ray volume. The vertical extent so far of the physical space is still spanning the corresponding vertical layer, i.e., $dz^{k-1/2}_{k+1/2}$. This is where the final multiplication by $(dz^{Z_u}_{Z_d}\dots)$ further collapses the physical space to only the vertical extent mutually common to both the ray, and the layer, i.e. $dz ({\textrm{ray }\cap\textrm{ layer}})$ where $\cap$ is the intersection operator. The last step is to then calculate the shear term for a given cell volume as a sum of contributions from all ray volumes populating that cell,
$$
S_{c,k}=\sum_{r}\frac{A_{c,k}(r)}{\rho_{k}}2m^2_{c,r}\tilde{m}^2_{c,k}(r)\frac{(\tilde{\hat{\omega}}(r)^2+f^2)_{c,k}}{(\tilde{\hat{\omega}}^2(r)\tilde{K}^2(r))_{c,k}}$$
%



%
%

\section*{Open Research Section}
This manuscript uses various models and vast amount of data all with various licenses and logistical limitations. To that end, the availability statements are as follows. The modified codes for Community Climate Icosahedral Nonhydrostatic Model ICON for atmosphere and climate as well as the Multi-Scale Gravity Wave Model MS-GWaM for gravity wave parameterization with the changes necessary to support the theoretical development of this paper fully implemented is publicly available from \cite{banerjee_2025_code}. The base ICON version used is 2.6.5-nwp1b with upper atmosphere extension which is described in \cite{borchert_upper-atmosphere_2019} and the base MSGWaM version used is 3D with pseudomomentum flux approximation which is described in \cite{voelker_ms-gwam_2024}. Furthermore, all the run scripts as well as post-processing scripts necessary to generate the data and plots from this manuscript can be publicly accessed from \cite{banerjee_2025_scripts}.

\acknowledgments
T.B., A.K. D.K., G.T.M, J.S. and U. A. thank the German Research Foundation (DFG) for partial support through CRC 301 “TPChange” (Project No. 428312742 and Projects B06 “Impact of small-scale dynamics on UTLS transport and mixing,” B07 “Impact of cirrus clouds on tropopause structure,” and Z03 “Joint model development and modelling synthesis”), and T.B., G.S.V. and U.A. thank DFG for support through the CRC 181 “Energy transfers in Atmosphere and Ocean” (Project No. 274762653 and Projects W01 “Gravity-wave parameterization for the atmosphere” and S02 “Improved Parameterizations and Numerics in Climate Models”). Work of Z. P. was partially supported by the Czech Science Agency under the project No. 23-04921M.

%
%

\bibliography{literature.bib}

%
%
%
%
%

\end{document}